\documentclass[twocolumn,english]{revtex4-1}
\usepackage[T1]{fontenc}
\usepackage[utf8]{inputenc}
\usepackage{amsmath}
\usepackage{amssymb}
\usepackage{graphicx}

\makeatletter
\usepackage[colorlinks = true,
            linkcolor =red,
            urlcolor  =magenta,
            citecolor = blue,
            anchorcolor = blue]{hyperref}

\makeatother

\usepackage{babel}
\begin{document}
\title{Effects of driven atomic ensemble on the output spectrum and entanglement
of optomechanical system}
\author{Burabigul Yakup\textsuperscript{1,2}, Yi-Fang Ren\textsuperscript{2}}
\author{Mamat Ali Bake\textsuperscript{1}}
\email{mabake@xju.edu.cn}

\author{Yusuf Turek\textsuperscript{2}}
\email{yusufu1984@hotmail.com}

\affiliation{\textsuperscript{1}Xinjiang Key Laboratory of Solid State Physics
and Devices, School of Physics Science and Technology, Urumqi 830017,
China}
\affiliation{\textsuperscript{2}School of Physics, Liaoning University, Shenyang,
Liaoning 110036, China}
\begin{abstract}
This paper considers an indirect driving model of a cavity QED system
in which the left cavity wall consists of a large ensemble of two-level
atoms driven by a classical laser field at a specific resonant frequency,
inducing an effective drive for the optomechanical system. We investigate
the effects of the atomic ensemble on the output intensity squeezing
spectrum and the entanglement between the optical and mechanical modes.
Our results show that both the coupling between the atomic ensemble
and the cavity mode and the excitation level of the atomic ensemble
significantly influences the output spectrum and the entanglement.
The theoretical model presented in this paper provides deeper insight
into the mechanisms governing correlations and squeezing spectra in
conventional optomechanical systems.
\end{abstract}
\maketitle

\section{Introduction }

Quantum entanglement and squeezing effects in nonclassical radiation
fields are fundamental aspects of quantum theory, playing a crucial
role in quantum information science and technology \citep{RevModPhys.80.517,RevModPhys.81.865,Nielsen_Chuang_2010,RevModPhys.84.777,RevModPhys.90.035005}.

Various physical systems have been explored to study the nonclassical
properties of quantum states, including ion traps \citep{PhysRevLett.75.4714,RevModPhys.75.281,RevModPhys.87.1419},
cavity QED \citep{RevModPhys.87.1379}, circuit QED \citep{RevModPhys.93.025005},
cavity optomechanics \citep{RevModPhys.86.1391,b1,b2}, and cavity
magnomechanics \citep{Zuo_2024}. Among these, cavity optomechanical
systems, which bridge optics, mechanics, and circuit electronics,
provide an excellent platform for investigating quantum correlations
and squeezing effects in optical and mechanical systems. Such investigations
enhances our understanding of the boundary between the classical and
quantum worlds \citep{RevModPhys.75.715,A23,RevModPhys.84.1655,RevModPhys.85.1083}.
Recent theoretical and experimental studies have focused on interactions
between macroscopic mechanical oscillators and cavity fields \citep{CHEN201797,Liao18,PhysRevLett.122.030402,EFTEKHARI2022127176,ZHANG2022127824,https://doi.org/10.1002/lpor.202301154},
revealing key quantum effects such as entanglement and squeezing.
In particular, quantum squeezing of mechanical modes plays a crucial
role in enhancing the precision of quantum measurements. Numerous
theoretical models have been proposed for generating mechanical mode
squeezing \citep{PhysRevA.83.033820,PhysRevA.88.063833,PhysRevA.79.063819,PhysRevA.82.033811}.
A typical cavity optomechanical system couples a movable mirror and
a cavity field via photon radiation pressure \citep{Xiao:14}. The
effective coupling between the mechanical and optical modes strongly
depends on the photon number inside the cavity, which can be significantly
enhanced when a two-level atomic ensemble is excited by a strong laser
field \citep{CHEN201797,Yu_2024}. Coherent control of optomechanical
systems through external laser pumping enables tunable coupling between
cavity modes and harmonic oscillators, leading to a range of interesting
applications \citep{PhysRevLett.98.030405,PhysRevLett.109.013603,4896}.
In such systems, quantum coherent operations can be realized by coupling
the mechanical oscillator to the optical cavity mode. For instance,
coherent coupling between mechanical and optical modes can be achieved
by introducing an auxiliary degree of freedom, such as superconducting
qubits, while ensuring that the coherence rate of energy exchange
exceeds the system’s decoherence rate \citep{PhysRevLett.99.213601,Pirandola_2006}.
The fundamental principle of cavity optomechanics is to leverage the
interaction between photons and mechanical oscillators to control
and measure mechanical motion \citep{Aspelmeyer2014}. Mechanical
oscillators are at the core of many high-precision experiments \citep{PhysRevD.100.066020}
and can exhibit exceptionally low dissipation \citep{verhagen2012quantum}.
Integrating an atomic ensemble into an optomechanical system enhances
nonlinear effects, significantly increasing optical nonreciprocity
compared to single-atom systems \citep{SONG201839}. When the atomic
ensemble is excited by a laser, its interaction with the cavity field
alters the frequency and intensity of the cavity field, which in turn
modifies the coupling strength between the cavity field and the mechanical
oscillator. This change in coupling strength can lead to quantum entanglement
between the optical and mechanical modes \citep{Vitali2007,Macovei2010QuantumEI,Aspelmeyer2014,Liao18}.
The atomic ensemble plays a key role in facilitating quantum information
transfer between subsystems through its interaction with the cavity
field. Additionally, the cavity field's output spectrum serves as
a powerful tool for probing the dynamics of optomechanical systems
\citep{1132,Kronwald2014,Pan_2020}, providing insights into the system's
quantum state, including entanglement, squeezing, and other nonclassical
correlations.

In this paper, we theoretically investigate the microscopic mechanism
governing the output spectrum of the cavity mode and the entanglement
phenomena between optical and mechanical modes in a standard cavity
optomechanical system. To achieve this, we employ an indirect driving
model of a cavity QED system, as introduced in Ref. \citep{PhysRevA.90.013836}.
In our model, the left wall of the cavity consists of a large ensemble
of local two-level atoms, which are driven first instead of directly
exciting the cavity mode. The right wall of the cavity is assumed
to be a movable mirror (mechanical resonator) that couples to the
cavity mode via radiation pressure induced by intracavity photon intensity.
The atomic ensemble, excited by an external classical driving field,
emits photons through spontaneous radiation, indirectly driving the
cavity mode. In this framework, the intracavity photon number is determined
by the excitation of the atomic ensemble. A higher photon number inside
the cavity results in stronger radiation pressure exerted on the mechanical
mode. Thus, the excitation level of the atomic ensemble directly influences
the coupling strength between the optical and mechanical modes in
the optomechanical system. Our results demonstrate that the excitation
level and the number of two-level atoms in the ensemble significantly
affect the system's squeezing and entanglement properties. Complete
squeezing can be achieved under moderate excitation of the atomic
condensate if its coupling with the cavity mode is sufficiently strong.
Regarding entanglement, we observe that increasing the number of atoms
in the ensemble and the excitation level via a classical driving field
leads to a significant enhancement in entanglement between the optical
and mechanical modes.

The paper is organized as follows. In Sec. \ref{sec:2}, we describe
our model using an effective Hamiltonian expressed in terms of collective
low-excitation operators of the atomic ensemble. In Sec. \ref{sec:3},
we derive the dynamical equations from the Hamiltonian and analyze
the fluctuation terms to obtain steady-state solutions. We then express
the output intensity spectrum based on input-output relations and
examine squeezing phenomena in the output intensity spectrum, focusing
on factors such as coupling intensity and atomic ensemble excitation
levels. In Sec. \ref{sec:5}, we introduce the orthogonal amplitude
and phase components of the cavity length and the corresponding noise
operators, reformulate the fluctuation terms of the Hamiltonian, and
quantify entanglement using logarithmic negativity. We analyze the
influence of coupling intensity, atomic ensemble excitation, and atom
number on the entanglement. Finally, Sec. \ref{sec:con} summarizes
our findings and conclusions.

\section{\label{sec:2} System Setup and Theoretical Framework }

This section presents the model and the fundamental theory underlying
our work. As shown in Fig. \ref{Fig:1}, the system under study is
a generic optomechanical system in which the left wall of the cavity
consists of a large ensemble of two-level atoms. A strong classical
external field with frequency $\omega_{f}$ drives the atomic ensemble.
As discussed in Ref. \citep{PhysRevA.90.013836}, this type of indirect
quantum driving can also excite the cavity modes. We express the model
Hamiltonian as follows (assuming $\hbar=1$):

\begin{figure}
\includegraphics[width=8cm]{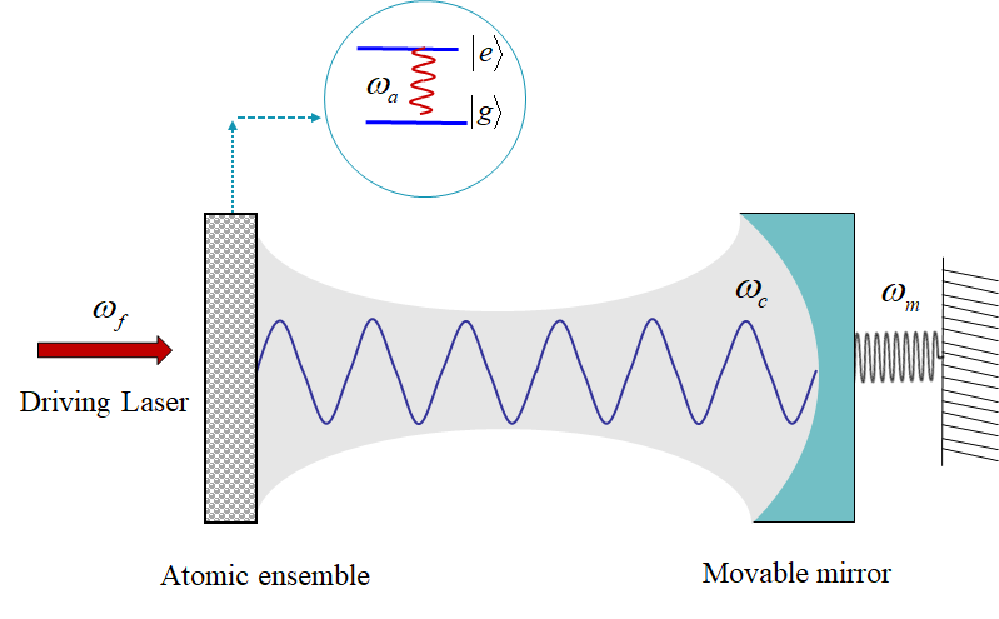}\caption{\label{Fig:1} Schematic of the system. The left wall of the cavity
consists of an ensemble of $N$ identical two-level atoms, each with
an energy difference of $\omega_{a}$. The atomic ensemble is driven
by a strong classical external field with frequency $\omega_{f}$,
which induces a driving effect on the single-mode cavity field.}
\end{figure}

\begin{align}
H & =\omega_{c}c^{\dagger}c+\frac{\omega_{a}}{2}\sum_{i=1}^{N}\sigma_{z}^{(i)}+\frac{\omega_{m}}{2}(p^{2}+x^{2})\nonumber \\
- & G_{0}c^{\dagger}cx+\sum_{i=1}^{N}[(gc+\Omega e^{-i\omega_{f}t})\sigma_{+}^{(i)}+H.c.].\label{eq:1}
\end{align}
Here, the first term represents the free Hamiltonian of the cavity,
where $c$ $(c^{\dagger})$ is the annihilation (creation) operator
of the single-mode cavity field with frequency $\omega_{f}$. The
second term corresponds to the free Hamiltonian of the two-level atomic
ensemble. The atomic operators are given by $\sigma_{z}^{(i)}=\vert e_{i}\rangle\langle e_{i}\vert-\vert g_{i}\rangle\langle g_{i}\vert$,
$\sigma_{+}^{(i)}=\vert e_{i}\rangle\langle g_{i}\vert$ and $\sigma_{-}^{(i)}=\vert g_{i}\rangle\langle e_{i}\vert$,
which are the Pauli matrices for the $i$-th atom. We assume that
the atomic ensemble consists of $N$ identical two-level atoms, all
sharing the same ground state $\vert g\rangle$ and excited state
$\vert e\rangle$, with an energy level spacing of $\omega_{a}$.
Each atom can therefore be described using the pseudo-spin-$1/2$
operators $\sigma_{z}$, $\sigma_{+}$, and $\sigma_{-}$, which satisfy
the commutation relations $[\sigma_{+},\sigma_{-}]=\sigma_{z}$ and
$[\sigma_{z},\sigma_{\pm}]=\pm2\sigma_{\pm}$. The third term describes
the Hamiltonian of the mechanical mode, characterized by a resonance
frequency $\omega_{m}$ and an effective mass $m$. The position $x$
and momentum $p$ operators of the mechanical oscillator satisfy the
canonical commutation relation $[x,p]=i\hbar$. The fourth term represents
the coupling between the mechanical oscillator and the cavity mode,
where the optomechanical coupling strength is given by $G_{0}=\frac{\omega_{c}}{L}\sqrt{\frac{\hbar}{m\omega_{m}}}$.
The last two terms describe the interaction Hamiltonian governing
the coupling between the atomic ensemble and the single-mode cavity
field, as well as the coupling between the atomic ensemble and the
external driving field with frequency $\omega_{f}$. The atomic ensemble,
which effectively acts as the left wall of the cavity, is arranged
in a thin layer with a negligible width compared to the wavelength
of the cavity field. Therefore, we assume all atoms coupled to the
cavity field with the same strength $g$ \citep{PhysRevLett.91.147903}.
Similarly, if the wavelength of the external driving field is smaller
than the characteristic size of the atomic ensemble, we can assume
that each atom couples to the external driving field with the same
strength $\Omega$ \citep{PhysRevA.90.013836}.

To simplify the Hamiltonian model, we apply the Holstein-Primakoff
(H-P) transformation \citep{PhysRev.58.1098} to the collective atomic
operators:
\begin{equation}
\sum_{i=1}^{N}\sigma_{+}^{(i)}=B^{\dagger}\sqrt{N-B^{\dagger}B},\label{eq:2-1}
\end{equation}
\begin{equation}
\sum_{i=1}^{N}\sigma_{-}^{(i)}=\sqrt{N-B^{\dagger}B}B,\label{eq:3-2}
\end{equation}
and
\begin{equation}
\sum_{i=1}^{N}\sigma_{z}^{(i)}=2B^{\dagger}B-N.\label{eq:4-1}
\end{equation}
Here, $B$ ($B^{\dagger}$) represents the collective atomic annihilation
(creation) operator. In the low-excitation regime, i.e., when $\langle B^{\dagger}B\rangle\ll N$,
the operators $B$ and $B^{\dagger}$ approximately satisfy the standard
bosonic commutation relation, $[B,B^{\dagger}]\approx1$. Under this
approximation, the transformation simplifies further as $\sum_{i=1}^{N}\sigma_{+}^{(i)}\approx B^{\dagger}\sqrt{N}$
and $\sum_{i=1}^{N}\sigma_{-}^{(i)}\approx\sqrt{N}B$ \citep{PhysRevB.68.134301,PhysRevB.71.205314}.

However, when the low-excitation condition for the atomic ensemble
is not met, higher-order terms in the Holstein-Primakoff transformation
must be included. In this work, we are particularly interested in
the high-excitation effects on the atomic ensemble. Therefore, when
considering the first-order expansion term, the Holstein-Primakoff
transformation modifies as follows:
\begin{equation}
\sum_{i=1}^{N}\sigma_{+}^{(i)}\approx B^{\dagger}\sqrt{N}(1-\frac{B^{\dagger}B}{2N}),
\end{equation}

\begin{equation}
\sum_{i=1}^{N}\sigma_{-}^{(i)}\approx B\sqrt{N}(1-\frac{B^{\dagger}B}{2N}),
\end{equation}
and
\begin{equation}
\sum_{i=1}^{N}\sigma_{z}^{(i)}=B^{\dagger}B-\frac{N}{2}.
\end{equation}
Using these relations, in the interaction picture \citep{CHEN201797,PhysRevA.90.013836}
with respect to $H_{0}=\omega_{f}(B^{\dagger}B+c^{\dagger}c)$, we
can rewrite our model Hamiltonian, Eq. (\ref{eq:1}), in terms of
the atomic collective operators $B$ and $B^{\dagger}$ as
\begin{align}
H_{I} & =\triangle_{c}c^{\dagger}c+\triangle_{a}B^{\dagger}B+\frac{\omega_{m}}{2}(p^{2}+x^{2})\label{eq:2}\\
 & -G_{0}c^{\dagger}cx+[(Gc+\chi)B^{\dagger}(1-\frac{B^{\dagger}B}{2N})+H.c],\nonumber
\end{align}
where $\triangle_{c}=\omega_{c}-\omega_{f}$ is the detuning between
the single-mode cavity and the external driving field, and $\triangle_{a}=\omega_{a}-\omega_{f}$
is the detuning between the two-level atom and the external driving
field. The parameters are defined as $G=g\sqrt{N}$, $\chi=\Omega\sqrt{N}$,
and $G_{0}=\frac{\omega_{c}}{L}\sqrt{\frac{\hbar}{m\omega_{m}}}$.
In the derivation of Eq. (\ref{eq:2}), we have neglected a constant
term $\frac{N\omega_{a}}{2}$, as it has no effect on our results
in this context.

\section{\label{sec:3} Quantum Dynamics and Output Field Spectrum}

We first calculate the quantum dynamical evolution of our hybrid optomechanical
system using the Heisenberg-Langevin equations of motion for the operators
$c$, $B$, $x$, and $p$. The quantum Langevin equations for the
system variables are derived from Eq. (\ref{eq:2}) as follows:

\begin{subequations}

\begin{align}
\dot{c} & =-(\kappa+i\triangle_{c})c-iGB+\frac{iG}{2N}B^{\dagger}B^{2}\label{eq:3-1}\\
 & +iG_{0}cx+\sqrt{2\kappa}c_{in}(t),\nonumber \\
\dot{B} & =-(\gamma_{a}+i\triangle_{a})B-iGc+(iGc+i\chi)\frac{B^{\dagger}B}{N}\label{eq:3}\\
 & +(iGc^{\dagger}+i\chi)\frac{B^{2}}{2N}-i\chi+\sqrt{2\gamma_{a}}B_{in}(t),\nonumber \\
\dot{x} & =\omega_{m}p,\label{eq:3-3}\\
\dot{p} & =-\omega_{m}x+G_{0}c^{\dagger}c-\gamma_{m}p+\xi(t).\label{eq:3-4}
\end{align}

\end{subequations}Here, $\kappa$ is the decay rate of the cavity,
$\gamma_{a}$ is the decay rate of the collective mode $B$, and $\gamma_{m}$
is the damping rate of the mechanical resonator. The operators $c_{in}(t)$
and $b_{in}(t)$ describe the quantum vacuum fluctuation of the cavity
field and the atomic ensemble, respectively, with zero mean values,
i.e., $\langle b_{in}(t)\rangle=\langle c_{in}(t)\rangle=0$. These
input vacuum noise operators satisfy the following nonzero fluctuation
relations:
\begin{align}
\langle c_{in}(t)c_{in}^{\dagger}(t^{\prime})\rangle & =[n(\omega_{c})+1]\delta(t-t^{\prime}),\label{eq:13-1}\\
\langle b_{in}(t)b_{in}^{\dagger}(t^{\prime})\rangle & =[n(\omega_{b})+1]\delta(t-t^{\prime}),\label{eq:13-2}
\end{align}
 where $n(\omega_{r})=[\exp(\frac{\hbar\omega_{r}}{\kappa_{B}T})-1]^{-1}$
($r=b,c$) represents the average thermal excitation number of the
cavity mode and the atomic collective mode at temperature $T$. $\xi(t)$
denotes the Brownian stochastic force with a zero mean value, and
its correlation function \citep{PhysRevA.63.023812} is given by
\begin{equation}
\langle\xi(t)\xi(t^{\prime})\rangle=\frac{\gamma_{m}}{\omega_{m}}\int\frac{d\omega}{2\pi}e^{-i\omega(t-t^{\prime})}\omega[1+\coth(\frac{\hbar\omega}{2\kappa_{B}T})],\label{eq:4}
\end{equation}
where $\kappa_{B}$ is the Boltzmann constant, and $T$ is the temperature
of the reservoir associated with the movable mirror. The steady-state
mean values of $c$, $B$, $x$, and $p$ can be obtained by setting
the time derivatives in Eqs. (\ref{eq:3-1}-\ref{eq:3-4}) to zero.
These values can be expressed as:
\begin{align}
P_{s} & =0,\\
x_{s} & =\frac{G_{0}\left|c_{s}\right|^{2}}{\omega_{m}},
\end{align}
and
\begin{equation}
c_{s}=\frac{-iG\sqrt{N}\frac{B_{s}}{\sqrt{N}}(1-\frac{1}{2}\left|\frac{B_{s}}{\sqrt{N}}\right|^{2})}{\kappa+i\triangle},\label{eq:18}
\end{equation}
respectively. Here, $\triangle=\triangle_{c}-G_{0}x_{s}$, and $\frac{B_{s}}{\sqrt{N}}$
is determined by the following equation:
\begin{equation}
-2(\triangle_{r}-i\gamma_{r})\frac{B_{s}}{\sqrt{N}}+2\frac{\left|B_{s}\right|^{2}}{N}+\frac{B_{s}^{2}}{N}-2=0\label{eq:6}
\end{equation}
with $\triangle_{r}=\frac{1}{\Omega}[\triangle_{a}-\frac{G^{2}\triangle}{\kappa^{2}+\triangle^{2}}(1-2\left|\frac{B_{s}}{\sqrt{N}}\right|^{2})]$
and $\gamma_{r}=\frac{1}{\Omega}[\gamma_{a}+\frac{G^{2}\triangle}{\kappa^{2}+\triangle^{2}}(1-\left|\frac{B_{s}}{\sqrt{N}}\right|^{2})]$.
We note that $\triangle_{r}$ and $\gamma_{r}$ characterize the effective
detuning and decay rate of the atomic ensemble \citep{CHEN201797,PhysRevA.93.023841}.
From Eq. (\ref{eq:18}), it is evident that the steady-state values
of the cavity field strongly depend on the higher-order excitation
$\vert B_{s}/\sqrt{N}\vert{}^{2}$ of the atomic ensemble. Consequently,
the effective optomechanical coupling is influenced by the higher-order
excitation, which affects the system dynamics and the entanglement
between the cavity mode and the mechanical oscillator.

Next, we consider fluctuations in the system around the steady-state
solution by examining the case where the photon number inside the
cavity is large, i.e., $\left|c_{s}\right|^{2}\gg1$. In this regime,
the operators can be expressed as $O=O_{s}+\delta O$, where $O=x,p,c,B$.
Substituting these expressions into Eqs. (\ref{eq:3-1})-(\ref{eq:3-4})
and retaining only first-order terms in the small quantities $\delta O$,
the quantum Langevin equations for the fluctuations can be written
as: \begin{subequations}

\begin{align}
\delta\dot{c} & =-(\kappa+i\triangle_{c})\delta c-iG_{2}\delta B+iG_{3}\delta B^{\dagger}\nonumber \\
 & +iG_{0}c_{s}\delta x+\sqrt{2\kappa}c_{in}(t),\label{eq:20}\\
\delta\dot{B} & =-(\gamma_{a}+i\triangle_{a}^{\prime})\delta B-iG_{2}\delta c+iG_{1}\delta B^{\dagger}\nonumber \\
 & +iG_{3}\delta c^{\dagger}+\sqrt{2\gamma_{a}}b_{in}(t),\label{eq:21}\\
\delta\dot{p} & =-\omega_{m}\delta x+G_{0}c_{s}\delta c^{\dagger}+G_{0}c_{s}\delta c-\gamma_{m}\delta p+\xi,\label{eq:22}\\
\delta\dot{x} & =\omega_{m}\delta p,\label{eq:23}
\end{align}
\end{subequations}where \begin{subequations}

\begin{align}
\triangle_{a}^{\prime} & =\triangle_{a}-2\frac{G}{\sqrt{N}}\Re[c_{s}^{\ast}(\frac{B_{s}}{\sqrt{N}})]-\frac{2\chi}{\sqrt{N}}\Re(\frac{B_{s}}{\sqrt{N}}),\label{eq:24}\\
G_{1} & =(Gc_{s}+\chi)\frac{B_{s}}{N},\label{eq:25}\\
G_{2} & =G\left(1-\frac{\left|B_{s}\right|^{2}}{N}\right),\label{eq:26}
\end{align}
\end{subequations}and $G_{3}=\frac{GB_{s}^{2}}{2N}$. From an experimental
perspective, it is convenient to transform the above equations into
the frequency domain using the Fourier transform (FT), defined by:
\begin{equation}
f(t)=\frac{1}{\sqrt{2\pi}}\int_{-\infty}^{+\infty}e^{-i\omega t}f(\omega)d\omega.\label{eq:12}
\end{equation}
Under the FT, the fluctuation correlation relations of the noise operators
are given by Eqs. (\ref{eq:13-1})-(\ref{eq:13-2}), and Eq. (\ref{eq:4})
transforms into: \begin{subequations}
\begin{align}
\langle c_{in}(\omega)c_{in}^{\dagger}(\omega^{\prime})\rangle & =[n(\omega_{c})+1]\delta(\omega-\omega^{\prime}),\label{eq:28}\\
\langle b_{in}\left(\omega\right)b_{in}^{\dagger}(\omega^{\prime})\rangle & =[n(\omega_{b})+1]\delta(\omega-\omega^{\prime}),\label{eq:30}
\end{align}
and
\begin{equation}
\langle\xi(\omega)\xi(\omega^{\prime})\rangle=\frac{\gamma_{m}}{\omega_{m}}\omega[1+\coth(\frac{\hbar\omega}{2\kappa_{B}})]\delta(\omega^{\prime}+\omega).\label{eq:30-1}
\end{equation}
 \end{subequations} We can express Eqs. (\ref{eq:20}-\ref{eq:23})
in matrix form as:
\begin{equation}
A(\omega)V(\omega)=Y(\omega),\label{eq:31}
\end{equation}
where

\begin{equation}
A(\omega)=\begin{pmatrix}\mu_{1} & 0 & iG_{2} & -iG_{3} & -iG_{0}c_{s} & 0\\
0 & \mu_{2} & iG_{3}^{*} & -iG_{2}^{*} & iG_{0}c_{s}^{*} & 0\\
iG_{2} & -iG_{3} & \nu_{1} & -iG_{1} & 0 & 0\\
iG_{3}^{*} & -iG_{2}^{*} & iG_{1}^{*} & \nu_{2} & 0 & 0\\
0 & 0 & 0 & 0 & i\omega & \omega_{m}\\
-G_{0}c_{s}^{*} & -G_{0}c_{s} & 0 & 0 & \omega_{m} & \gamma_{m}-i\omega
\end{pmatrix},\label{eq:8}
\end{equation}

\begin{equation}
V(\omega)=(\delta c(\omega),\delta c^{\dagger}(-\omega),\delta B(\omega),\delta B^{\dagger}(-\omega),\delta x(\omega),\delta p(\omega))^{T},\label{eq:32}
\end{equation}
 and

\begin{align}
Y(\omega) & =(\sqrt{2\kappa}c_{in}(\omega),\sqrt{2\kappa}c_{in}^{\dagger}(-\omega),\sqrt{2\gamma_{a}}b_{in}(\omega)\nonumber \\
, & \sqrt{2\gamma_{a}}b_{in}^{\dagger}(-\omega),0,\xi(\omega))^{T}\label{eq:33}
\end{align}
with $\mu_{1}=\kappa+i(\triangle-\omega)$, $\mu_{2}=\kappa-i(\triangle+\omega)$,
$\nu_{1}=\gamma_{a}+i(\triangle_{a}^{\prime}-\omega)$, and $\nu_{2}=\gamma_{a}-i(\triangle_{a}^{\prime}+\omega)$.
For the cavity field, the input-output relation is given by $c_{out}(t)=\sqrt{2k}c(t)-c_{in}(t)$
which, in the frequency domain, takes the form $c_{out}(\omega)=\sqrt{2k}c(\omega)-c_{in}(\omega)$.
The reason for applying the Fourier transform (FT) is that, in optical
experiments, it is more convenient to measure the relevant quantities
in the frequency domain rather than in the time domain. Next, we use
these quantities to obtain the squeezing spectrum of the output (transmitted)
field.

The output intensity squeezing spectrum of the field is defined as
\citep{Pan_2020,PhysRevA.46.4397}:
\begin{equation}
S_{out}\left(\omega\right)=\frac{1}{|\alpha_{out}|^{2}}\int d\omega^{\prime}\langle\delta I^{out}\left(\omega\right)\delta I^{out}\left(\omega^{\prime}\right)\rangle,\label{eq:34}
\end{equation}
where
\begin{equation}
\delta I_{out}\left(\omega\right)=\alpha_{out}^{*}\delta c_{out}\left(\omega\right)+\alpha_{out}\delta c_{out}^{\dagger}\left(-\omega\right)\label{eq:35}
\end{equation}
and $\delta c_{out}(\omega)=\sqrt{2\kappa}\delta c(\omega)-\delta c_{in}(\omega)$.
For the field quadrature operator $X_{\theta}^{out}$ we know that
$[X_{\theta}^{out},X_{\theta+\frac{\pi}{2}}^{out}]=2i$, thus, the
intensity squeezing occurs when $S_{out}(\omega)<1$. By using Eq.
(\ref{eq:31}) and the input-output relations, we can obtain $\delta c_{out}(\omega)$
as:
\begin{align}
\delta c_{out}(\omega) & =A(\omega)\delta c_{in}(\omega)+B(\omega)\delta c_{in}^{\dagger}(-\omega)+C(\omega)\delta B_{in}(\omega)\nonumber \\
+ & D(\omega)\delta B_{in}^{\dagger}(-\omega)+F(\omega)\xi(\omega).\label{eq:37}
\end{align}
The explicit expression of $A(\omega)$, $B(\omega)$, $C(\omega)$,
$D(\omega)$, and $F(\omega)$ are listed in Appendix \ref{sec:A1}.
By substituting the above-related quantities into Eq. (\ref{eq:34}),
we obtain the expression for the output intensity squeezing spectrum
of the cavity field, given by:
\begin{align}
S_{opt}(\omega) & =\left|A(\omega)+C(\omega)\right|^{2}+\left|B(-\omega)+D(-\omega)\right|^{2}\nonumber \\
+ & (\left|F(\omega)\right|^{2}+\left|F(-\omega)\right|^{2})\frac{\gamma_{m}}{\omega_{m}}\omega[-1+\coth(\frac{\hbar\omega}{2\kappa_{B}T})]\nonumber \\
- & 2\left|(A(\omega)+C(\omega))(B(-\omega)+D(-\omega))\right.\nonumber \\
+ & \left.F(\omega)F(-\omega)\frac{\gamma_{m}}{\omega_{m}}\omega[-1+\coth(\frac{\hbar\omega}{2\kappa_{B}T})]\right|.\label{eq:38}
\end{align}
It can be observed from the formula that the output spectrum of the
cavity field fluctuation term depends on the input vacuum noise and
the thermal noise of the mechanical oscillator. Compared to a classical
cavity optical force system, the coupling between the single-mode
cavity field and the mechanical oscillator also indirectly affects
the output spectrum of the cavity field fluctuation term.

Next, we focus on how the degree of squeezing depends on the coupling
strength between the cavity field and the atomic ensemble. We consider
an atomic ensemble with $N=10^{6}$ placed inside a cavity driven
by a laser. The cavity has a length of $L=10^{-3}$m and a decay rate
of $\kappa=2\pi\times2.5\times10^{6}$Hz. The decay rate of the atomic
ensemble is $\gamma_{a}=20\kappa$. The coupling between the cavity
mode and the mechanical resonator is given by $G_{0}=\frac{\omega_{c}}{L}\sqrt{\frac{\hbar}{m\omega_{m}}}$
where we take the mass of the mechanical resonator as $m=10^{-11}$
kg and its frequency as $\omega_{m}=2\pi\times4\times10^{7}$Hz.
\begin{figure}
\includegraphics[width=8cm]{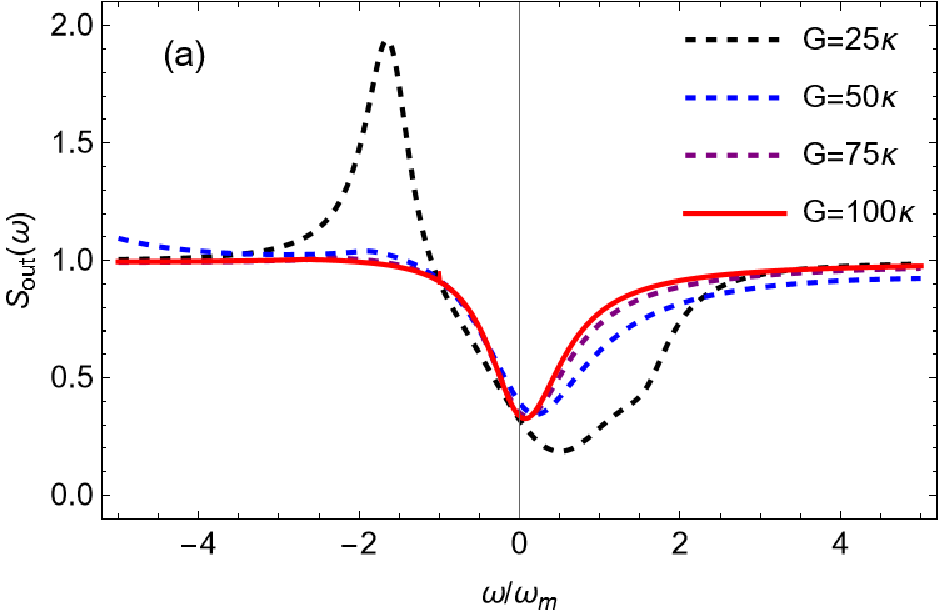}

\includegraphics[width=8cm]{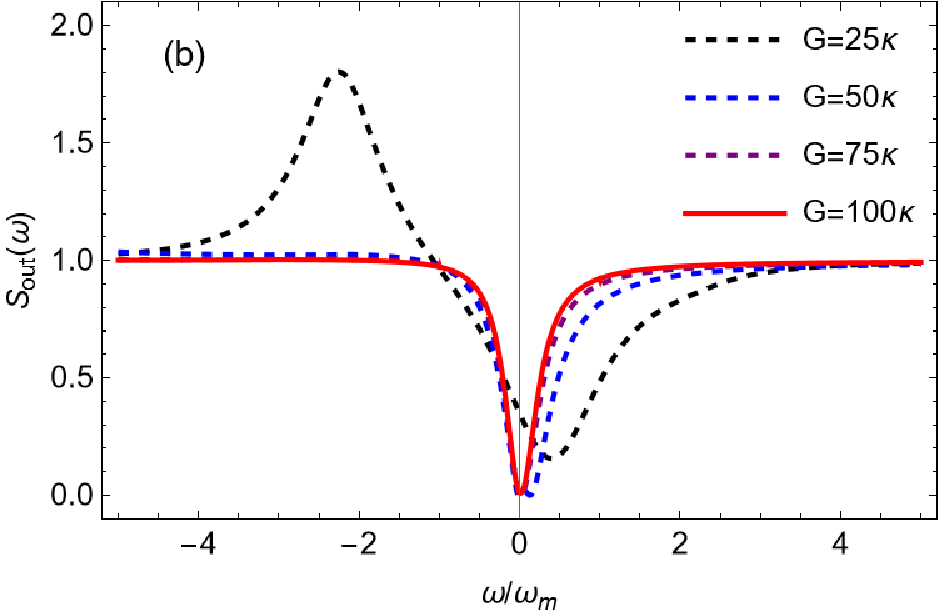}

\includegraphics[width=8cm]{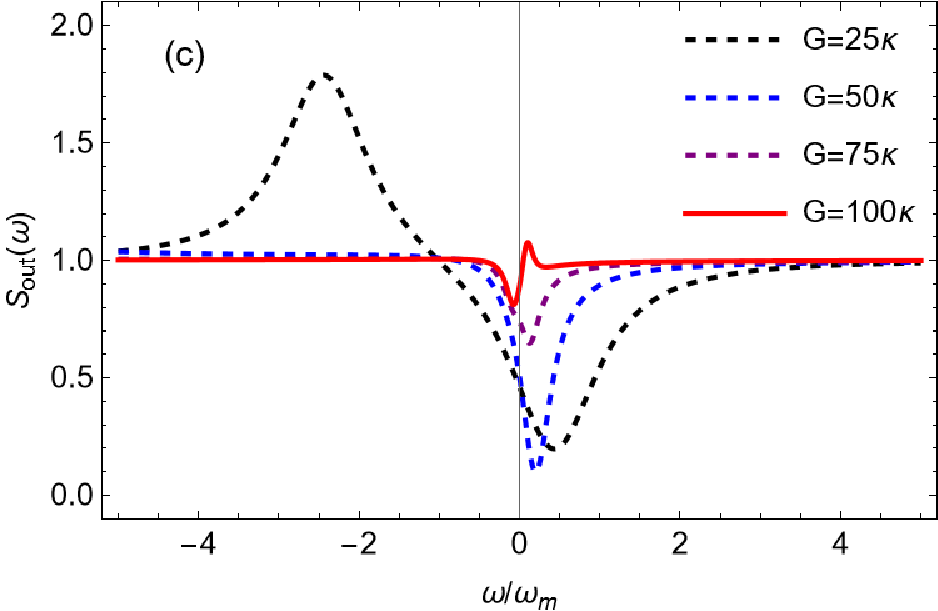}

\caption{\label{Fig:2} The output spectrum $S_{out}(\omega)$ as a function
of the normalized frequency $\omega/\omega_{m}$. The parameters used
are: $\omega_{m}=2\pi\times4\times10^{7}$Hz, $\kappa=2\pi\times2.5\times10^{6}$Hz,
$\gamma_{a}=20\kappa$, $\gamma_{m}=10^{-3}\omega_{m}$, $N=10^{7}$,
$\Delta=-\omega_{m}$, $L=1$mm, $m=10^{-13}$kg. (a) $\Delta_{r}=\gamma_{r}=1$,
(b) $\Delta_{r}=\gamma_{r}=2.5$, (c) $\Delta_{r}=\gamma_{r}=8$.}
\end{figure}

We now discuss the crucial role of atomic ensemble excitation in shaping
the properties of the output field. According to Eq. (\ref{eq:6}),
the excitation number $\mid B_{s}/\sqrt{N}\mid^{2}$ is determined
by the dimensionless effective detuning and decay rate $\Delta_{r}$
and $\gamma_{r}$, respectively. We consider three cases: $\Delta_{r}=\gamma_{r}=1$,
$\Delta_{r}=\gamma_{r}=2.5$ and $\Delta_{r}=\gamma_{r}=8$. The corresponding
excitation numbers are calculated as $\mid B_{s}/\sqrt{N}\mid^{2}=0.255$,
$\mid B_{s}/\sqrt{N}\mid^{2}=0.069$, and $\mid B_{s}/\sqrt{N}\mid^{2}=0.008$,
respectively.

In Fig. \ref{Fig:2} (a), we examine the higher excitation case for
$\Delta_{r}=\gamma_{r}=1$, which corresponds to $\mid B_{s}/\sqrt{N}\mid^{2}=0.255$.
We find that a squeezing effect occurs, and it increases with increasing
coupling strength $G$. In Fig. \ref{Fig:2} (b), the curves represent
the medium excitation case for $\Delta_{r}=\gamma_{r}=2.5$, with
a corresponding excitation number of $\mid B_{s}/\sqrt{N}\mid^{2}=0.069$.
Compared to Fig. \ref{Fig:2} (a), the width of the curves in Fig.
\ref{Fig:2} (b) becomes narrower, and the squeezing effect becomes
more pronounced with increasing coupling strength $G$, bringing the
system closer to the resonance point. As indicated in Fig. \ref{Fig:2}
(b), complete squeezing, corresponding to $S_{out}(\omega)=0$, can
be achieved for larger coupling strengths, i.e., $G=50\kappa,75\kappa,100\kappa$.

As shown in Fig. \ref{Fig:2} (c), in the lower excitation case where
$\Delta_{r}=\gamma_{r}=8$, the excitation number is reduced to $\mid B_{s}/\sqrt{N}\mid^{2}=0.008$.
Even in this regime, a squeezing effect is still observed, but with
a narrower width compared to the higher excitation cases {[}see Fig.
\ref{Fig:2} (a){]}. However, as the coupling strength increases,
the squeezing effect gradually diminishes. When the low-excitation
condition is slightly violated, the coupling between the cavity mode
and the atomic ensemble, $G$, is fundamentally determined by the
coupling coefficients $G_{1,2,3}$, which depend on the excitation
number of the atomic ensemble \citep{PhysRevA.93.023841}.

As mentioned earlier, the squeezing effect for a coupling strength
of $G=100\kappa$ is more pronounced in the $\Delta_{r}=\gamma_{r}=2.5$
case, which is an interesting result. Furthermore, it is important
to note that the coupling strength between the cavity mode and the
atomic ensemble is essential for achieving the squeezing effect. Our
analytic and numerical calculations show that when the coupling strength
is set to $G=0$, no squeezing is observed (the curve not plotted
in this paper). This provides an effective method for controlling
the degree of squeezing and optimizing the squeezing conditions.

\section{\label{sec:5} Effects of the Atomic Ensemble on Steady-State Entanglement}

In this section, we investigate the effects of the different excitation
limits of the atomic ensemble on the steady-state entanglement between
the cavity field and the mechanical resonator. To achieve this, we
begin our analysis with the linearized dynamics of the quantum fluctuations
in our system. The operators in Eq. (\ref{eq:3}) can be separated
into their steady-state expectation values at the fixed point (steady
state) and quantum fluctuations \citep{PhysRevA.86.063809}, i.e.,
$a=a_{s}+\delta a$, $(a=x,p,c,c^{\dagger})$. The constant steady-state
expectation values are given by Eqs. (\ref{eq:3-1})-(\ref{eq:3-4}).
By introducing the quadrature fluctuation operators $\delta X=(\delta c+\delta c^{\dagger})/\sqrt{2}$,
$\delta Y=(\delta c-\delta c^{\dagger})/\sqrt{2}i$, $\delta U=(\delta B+\delta B^{\dagger})/\sqrt{2}$,
$\delta V=(\delta B-\delta B^{\dagger})/\sqrt{2}i$ and the corresponding
Hermitian input noise operators $\delta X_{in}=(\delta c_{in}+\delta c_{in}^{\dagger})/\sqrt{2}$,
$\delta Y_{in}=(\delta c_{in}-\delta c_{in}^{\dagger})/\sqrt{2}i$,
$\delta u_{in}=(\delta b_{in}+\delta b_{in}^{\dagger})/\sqrt{2}$,
$\delta v_{in}=(\delta b-\delta b^{\dagger})/\sqrt{2}i$, we can rewrite
Eqs. (\ref{eq:3-1})-(\ref{eq:3-4}) as: :\begin{subequations}

\begin{align}
\delta\dot{x}\left(t\right) & =\omega_{m}\delta p,\\
\delta\dot{p}\left(t\right) & =-\omega_{m}\delta x-\gamma_{m}\delta p+G_{px}\delta X+G_{py}\delta Y+\xi,\\
\delta\dot{X}\left(t\right) & =-\kappa\delta X+\Delta\delta Y+G_{2}\delta V+G_{3\mu}\delta U+G_{3\nu}\delta V\nonumber \\
- & G_{py}\delta x+\sqrt{2\kappa}\delta X_{in},\\
\delta\dot{Y}\left(t\right) & =-\kappa\delta Y-\Delta\delta X-G_{2}\delta U+G_{3\nu}\delta U-G_{3\mu}\delta V\nonumber \\
+ & G_{px}\delta x+\sqrt{2\kappa}\delta Y_{in},\\
\delta\dot{U}\left(t\right) & =-\gamma_{a}\delta U+\triangle_{a}^{\prime}\delta V+G_{2}\delta Y+G_{\mu}\delta U+G_{\nu}\delta V\nonumber \\
+ & G_{3\mu}\delta X+G_{3\nu}\delta Y+\sqrt{2\gamma_{a}}\delta u_{in},\\
\delta\dot{V}\left(t\right) & =-\gamma_{a}\delta V-\triangle_{a}^{\prime}\delta U-G_{2}\delta X+G_{\nu}\delta U-G_{\mu}\delta V\nonumber \\
+ & G_{3\nu}\delta X-G_{3\mu}\delta Y+\sqrt{2\gamma_{a}}\delta v_{in}.
\end{align}

\end{subequations}

The linearized dynamical equations of the system can be written as
follows:
\begin{equation}
\dot{f}\left(t\right)=Jf\left(t\right)+n\left(t\right),\label{eq:13}
\end{equation}
where $f^{T}(t)=(\delta x(t),\delta p(t),\delta X(t),\delta Y(t),\delta U(t),\delta V(t))$
is the column vector of fluctuation operators, and $n(t)^{T}=(0,\xi,\sqrt{2\kappa}\delta X_{in},\sqrt{2\kappa}\delta Y_{in},\sqrt{2\gamma_{a}}\delta u_{in},\sqrt{2\gamma_{a}}\delta v_{in})$
is the corresponding noise vector. The matrix $J$ is the drift matrix,
given by:
\begin{equation}
J=\left(\begin{array}{cccccc}
0 & \omega_{m} & 0 & 0 & 0 & 0\\
-\omega_{m} & -\gamma_{m} & G_{px} & G_{py} & 0 & 0\\
-G_{py} & 0 & -\kappa & \Delta & G_{3\mu} & t_{1}\\
G_{px} & 0 & -\Delta & -\kappa & t_{2} & -G_{3\mu}\\
0 & 0 & G_{3\mu} & t_{3} & t_{4} & t_{5}\\
0 & 0 & t_{6} & -G_{3\mu} & t_{7} & t_{8}
\end{array}\right)\label{eq:14}
\end{equation}
where $t_{1}=G_{2}+G_{3\nu}$, $t_{2}=G_{3\nu}-G_{2}$, $t_{3}=G_{2}+G_{3\nu}$,
$t_{4}=G_{\mu}-\gamma_{a}$, $t_{5}=G_{\nu}+\triangle_{a}^{\prime}$,
$t_{6}=G_{3\nu}-G_{2}$, $t_{7}=G_{\nu}-\triangle_{a}^{\prime}$,
$t_{8}=-\gamma_{a}-G_{\mu}$, $G_{px}=2G_{0}Re[c_{s}]/\sqrt{2}$,
and $G_{py}=-i2G_{0}Im[c_{s}]/\sqrt{2}$ are the effective optomechanical
coupling, and where $G_{\mu}=i\left(G_{1}-G_{1}^{*}\right)/\sqrt{2}$,
$G_{\nu}=2Re[G_{1}]/\sqrt{2}$, $G_{\mu}=2iIm[G_{2\mu}]/\sqrt{2}$,
and $G_{\nu}=2Re[G_{2\nu}]/\sqrt{2}$. From Eq . (\ref{eq:14}), we
see that the higher-order coupling coefficients $G_{\nu}$, $G_{\mu}$,
$G_{3\nu}$, and $G_{3\mu}$ can strongly influence the fluctuation
dynamics of the hybrid optomechanical system and the generation of
steady-state entanglement between the two coupled bosonic modes. The
steady state of the hybrid system is unique when the solutions to
Eq. (\ref{eq:13}) are stable, i.e., when the real parts of all eigenvalues
of the matrix $J$ are negative. The stability condition can be reformulated
using the Routh-Hurwitz criteria \citep{PhysRevA.35.5288,CHEN201797}.

Due to the Gaussian nature of the noise term in Eq. (\ref{eq:13})
and the linearized dynamic equations, the quantum steady state for
the fluctuations is a zero-mean bipartite Gaussian state, fully characterized
by its $6\times6$ correlation matrix: $V_{ij}=(\langle u_{i}(\infty)u_{j}(\infty)+u_{j}(\infty)u_{i}(\infty)\rangle)/2$
where $u^{T}(\infty)=(\delta x(\infty),\delta p(\infty),\delta X(\infty),\delta Y(\infty))$
is the vector of continuous variable fluctuation operators at the
steady state $(t\rightarrow\infty)$. The Brownian noise $\xi(t)$
is not $\delta$-correlated and therefore does not describe a Markovian
process. The steady-state correlation matrix $V$ can be obtained
by solving the linearized Langevin equation (\ref{eq:13}), which
satisfies the Lyapunov equation: $JV+VJ^{T}=-D$, where $D=diag[0,\gamma_{m}(2n+1),\kappa,\kappa,\gamma_{a},\gamma_{a}]$.

Furthermore, to verify the generation of optomechanical entanglement
due to the optomechanical coupling, we analyze the bipartite entanglement
properties between the cavity field and the mechanical mode, which
are quantified using the logarithmic negativity $E_{N}$.
\begin{equation}
E_{N}=\max[0,-\ln2\nu].\label{eq:15}
\end{equation}
Here, $\nu$ is the smallest symplectic eigenvalue of the partially
transposed correlation matrix associated with the selected bipartite
subsystem. This matrix is obtained by tracing out the remaining degrees
of freedom, and the corresponding reduced correlation matrix can be
represented in the $2\times2$ block form as: $V=\left(\begin{array}{cc}
A & C\\
C^{T} & B
\end{array}\right)$ \citep{Chen:17}.

In this manner, the smallest symplectic eigenvalue is given by:
\[
\nu=\frac{1}{\sqrt{2}}\sqrt{\Sigma(V)-[\Sigma(V)^{2}-4\det V]^{\frac{1}{2}}}
\]
with $\sum(V)=detA+detB-2\,detC$.

We now numerically evaluate the steady-state entanglement between
the mechanical and optical modes. The stationary entanglement of the
optomechanical system is well-defined in the regime where the system
remains in a single stable state \citep{Zhang_2013}.

\begin{figure}
\includegraphics[width=8cm]{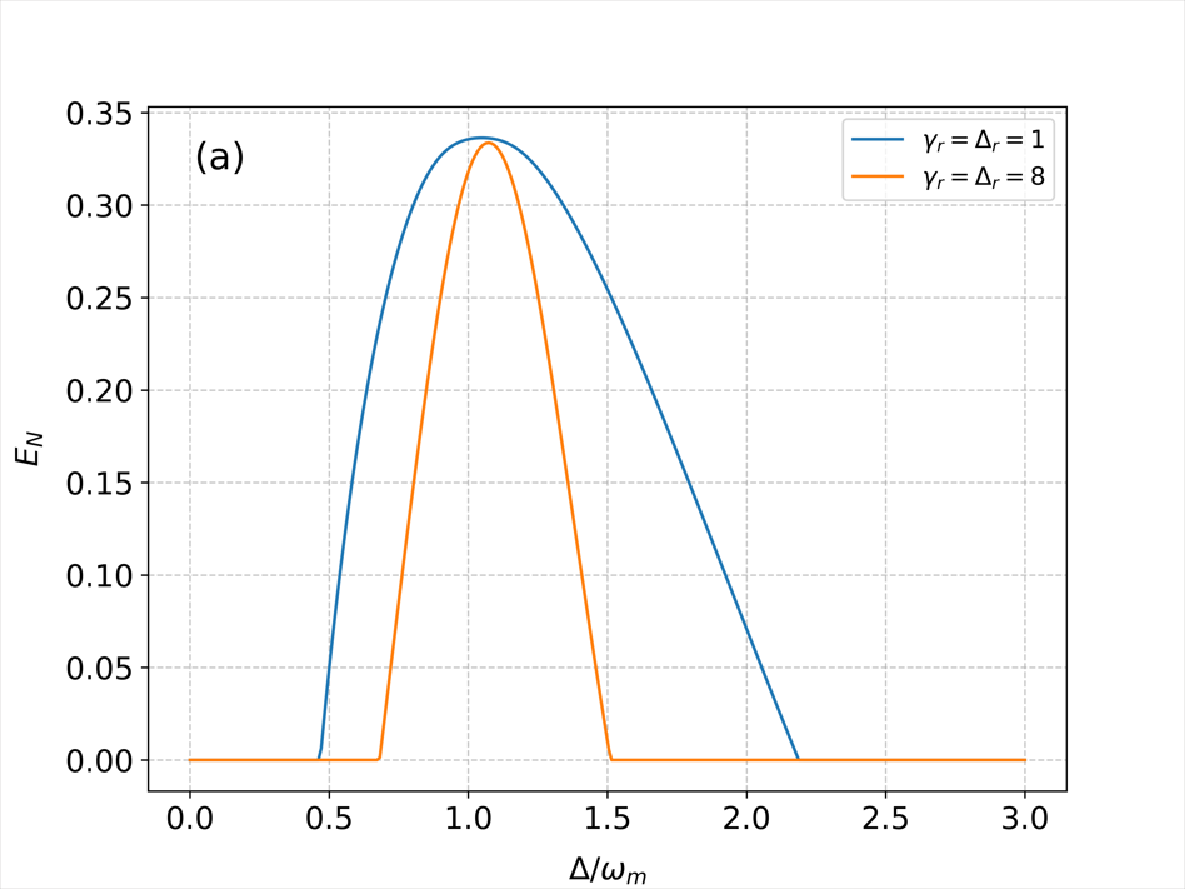}

\includegraphics[width=8cm]{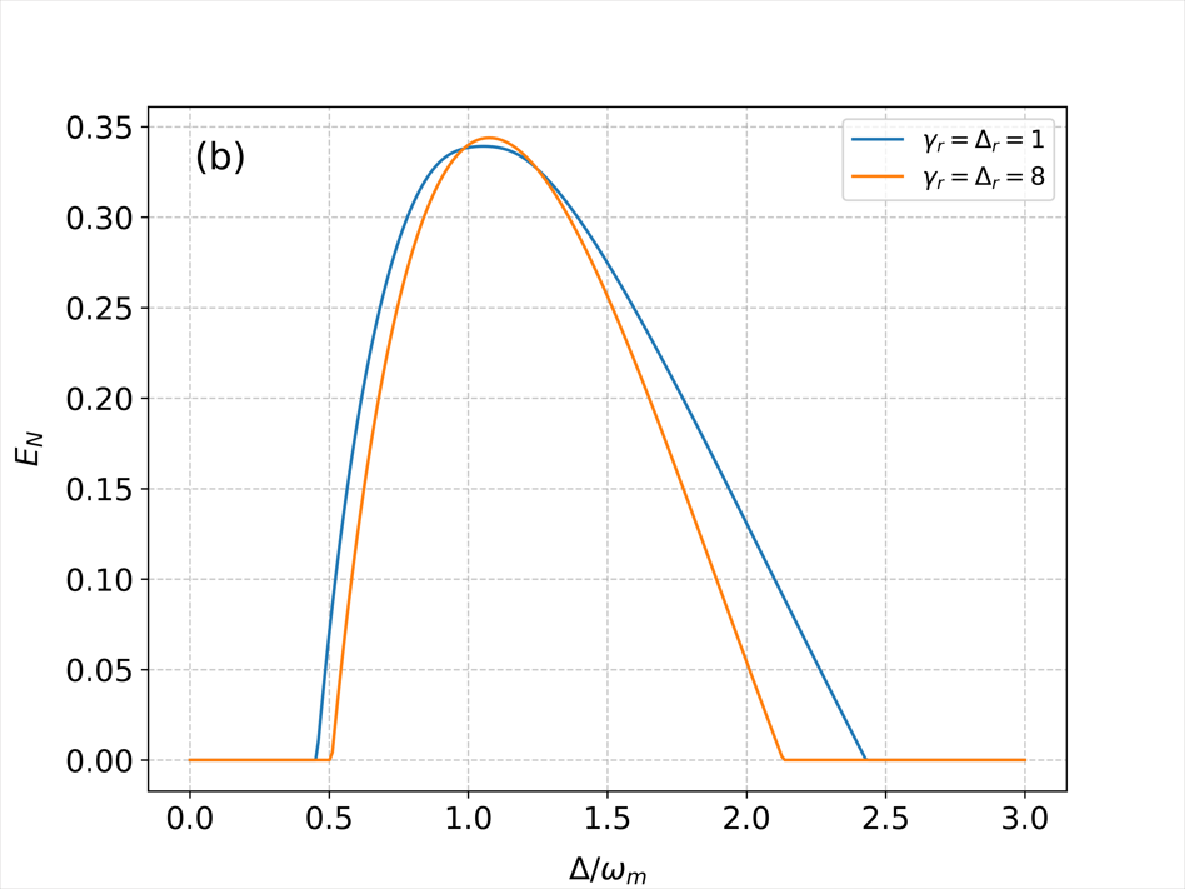}

\caption{\label{fig:Fig3} Plot of the steady-state entanglement $E_{N}$ as
a function of the dimensionless effective detuning $\Delta/\omega_{m}$
for different atomic excitation cases. (a) $G=25\kappa$, (b) $G=100\kappa$.
Here, we take $N=10^{7}$, and other parameter values are the same
as in Fig. \ref{Fig:2}.}
\end{figure}

Fig. \ref{fig:Fig3} shows the steady-state entanglement between the
cavity mode and the mechanical oscillator, measured by the logarithmic
negativity $E_{N}$, as a function of the dimensionless effective
detuning $\Delta/\omega_{m}$. The cases considered are $B_{s}/\sqrt{N}=-0.411-0.291i$
and $B_{s}/\sqrt{N}=-0.062-0.061i$, with the corresponding excitation
numbers calculated as $\left|B_{s}/\sqrt{N}\right|^{2}=0.255$ and
$\left|B_{s}/\sqrt{N}\right|^{2}=0.008$, respectively. These two
cases correspond to the high and low-excitation regimes of the atomic
ensemble.

As shown in Fig. \ref{fig:Fig3} (a), optomechanical entanglement
between the cavity mode and the mechanical oscillator can be achieved
in the region of positive detuning. In the case of weak coupling strength
$G=25\kappa$, the entanglement for the high-excitation state ($\Delta_{r}=\gamma_{r}=1$)
is larger than that of the low-excitation state ($\Delta_{r}=\gamma_{r}=8$).
Additionally, for $G=25\kappa$, the range of effective detuning $\Delta/\omega_{m}$
where entanglement occurs is relatively narrower in the lower-excitation
case than in the higher-excitation case. Notably, an optimal effective
detuning maximizes the optomechanical entanglement. For instance,
the peak entanglement, $E_{N}^{\max}\simeq0.34$, occurs at $\Delta\simeq1.22\omega_{m}$.

In Fig. \ref{fig:Fig3} (b), with a strong coupling strength of $G=100\kappa$,
the peak of the curve for the low-excitation state is slightly higher
than that of the high-excitation state. Additionally, the range of
effective detuning $\Delta/\omega_{m}$ increases, while the entanglement
$E_{N}$ decreases gradually with increasing coupling strength. This
effect occurs because, when the low-excitation condition breaks, a
strong cavity field driving and a relatively small atom-field detuning
can enhance the optomechanical coupling between the mechanical and
optical modes, thereby facilitating the generation of steady-state
entanglement \citep{CHEN201797}. Consequently, the degree of entanglement
increases slightly with increasing coupling strength.
\begin{figure}
\includegraphics[width=8cm]{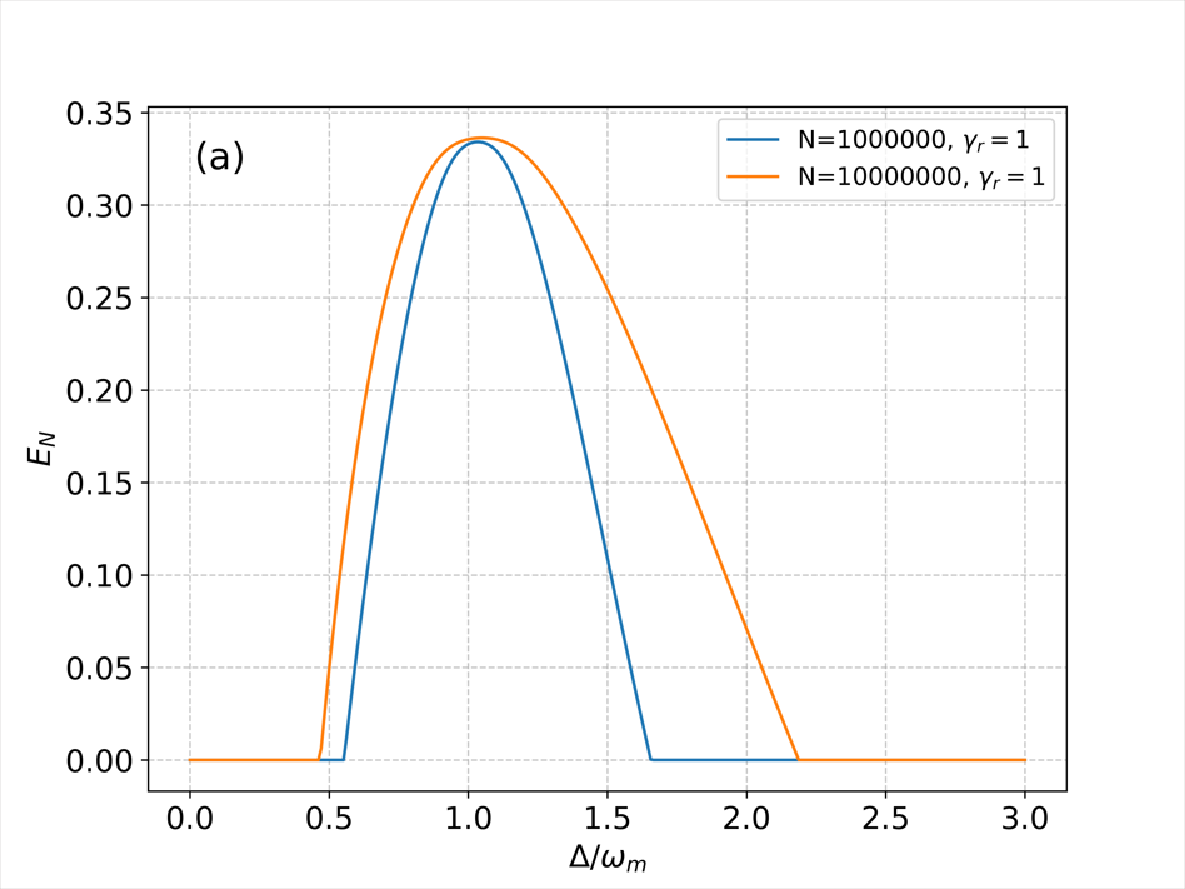}

\includegraphics[width=8cm]{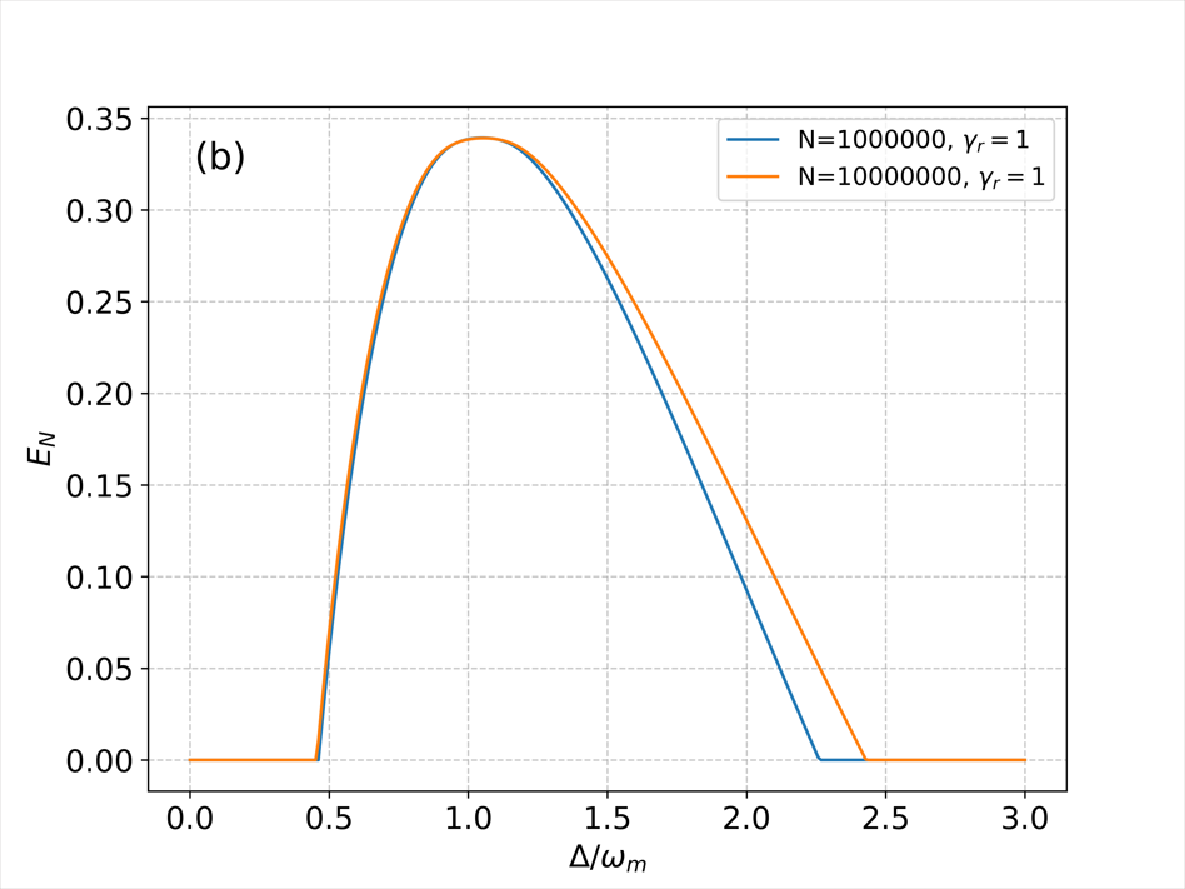}

\caption{\label{fig:Fig4} Plot of the steady-state entanglement $E_{N}$ as
a function of the dimensionless effective detuning $\Delta/\omega_{m}$
for different atomic ensembles sizes: $N=10^{6}$ and $N=10^{7}$,
in the case of $\Delta_{r}=\gamma_{r}=1$. The results are shown for
(a) $G=25\kappa$ and (b) $G=100\kappa$. Other parameters are the
same as in Fig. \ref{Fig:2}.}
\end{figure}

We also discuss the importance of atomic ensemble excitation in generating
mechanical oscillator entanglement. In Fig. \ref{fig:Fig4}, we show
the steady-state entanglement $E_{N}$ as a function of the dimensionless
effective detuning $\Delta/\omega_{m}$ for different atomic ensemble
sizes. We consider the case of $\Delta_{r}=\gamma_{r}=1$, where the
corresponding excitation number is calculated as $\left|B_{s}/\sqrt{N}\right|^{2}=0.255$,
which corresponds to the presence of higher-order atomic excitation.
As seen in Fig. \ref{fig:Fig4}, the number of atoms in the ensemble
influences the optomechanical entanglement and its maximum value.
In Fig. \ref{fig:Fig4} (a), under weak coupling strength $G=25\kappa$,
the maximum entanglement values for $N=10^{6}$ and $N=10^{7}$ are
the same. However, for $N=10^{6}$, the entanglement region is narrower
than for $N=10^{7}$. In both cases, the degree of entanglement diminishes
with increasing effective detuning $\Delta$.

Furthermore, as shown in Fig. \ref{fig:Fig4} (b), under strong coupling
strength $G=100\kappa$, the peaks of the two curves coincide, and
the range of effective detuning $\Delta/\omega_{m}$ increases for
both $N=10^{6}$ and $N=10^{7}$. In other words, in a highly excited
atomic ensemble, increasing the coupling strength causes the entanglement
curve for $N=10^{6}$ to approach that of $N=10^{7}$. Additionally,
the maximum entanglement increases with increasing coupling strength.
When the maximum entanglement is reached, the optimal effective detuning
approaches the mechanical frequency of the oscillator, i.e., $\Delta\approx\omega_{m}$.

In summary, the entanglement between the mechanical and optical modes
is closely related to the excitation level of the atomic ensemble,
which acts as the cavity wall, and its degree depends on the photon
intensity inside the cavity generated by the indirect driving. When
the atomic ensemble is highly excited, the degree of entanglement
is generally superior to that in the low-excitation case. By increasing
the coupling strength $G$ and the number of atoms $N$ in the ensemble,
the range of effective detuning $\Delta$ over which entanglement
occurs can be significantly broadened.

\section{\label{sec:con}Conclusion}

In this paper, we have investigated the entanglement and output spectrum
of an optomechanical system using the microscopic mechanism of external
driving for a single-mode cavity field based on an indirect driving
model. In this simplified model, one wall of the Fabry-Pérot cavity
is assumed to be an ensemble of local two-level systems. An external
driving field is applied to the atomic ensemble exciting the atoms
of the left wall. The excitation of the atomic ensemble provides indirect
driving to the cavity mode and induces coupling between the optical
cavity mode and the mechanical resonator mode.

From the output intensity squeezing spectrum of the single-mode cavity
field, we found that squeezing of the output field occurs for different
excitation levels of the atomic ensemble when the driving strength
increases. Notably, when the atomic ensemble is at a medium excitation
level, complete squeezing of the output cavity mode can be achieved
for a sufficiently large coupling strength between the atomic ensemble
and the cavity mode. We also found that maximum squeezing occurs near
the resonance point. Furthermore, we investigated the effects of the
atomic ensemble on the entanglement between the optical and mechanical
modes of the system and characterized the degree of entanglement in
terms of logarithmic negativity. The analytical and numerical results
showed that the degree of entanglement is not only affected by the
coupling strength $G$ but is also closely related to the number of
atoms in the ensemble $N$. Specifically, for a fixed number of atoms
forming the left cavity wall, a higher excitation level results in
stronger entanglement compared to a lower excitation level when the
coupling strength $G$ increases. However, in the lower excitation
case, significant entanglement can still be achieved, and the entanglement
regions and maximum entanglement points are nearly the same as those
in the higher excitation case if the coupling strength between the
atomic ensemble and the optical cavity mode is significantly increased.

In analyzing the effects of different numbers of atoms on the degree
of entanglement between the optical and mechanical modes, we found
that for a higher excitation, the entanglement in the system is proportional
to the number of atoms $N$ and the coupling strength $G$. In other
words, significant strong entanglement between the optical and mechanical
modes of the system can be achieved if there is strong coupling strength
$G$ and the left cavity wall consists of a large number of local
two-level atoms. We also observed that the maximum entanglement points
in our system always occur near $\triangle\approx\omega_{m}$. In
all cases, the degree of entanglement diminishes when the effective
detuning is either very small or very large. In summary, a higher
coupling strength $G$ enhances entanglement, as strong coupling facilitates
quantum information transfer between atoms more effectively, thereby
enhancing atomic correlations.

The model setup and theoretical results presented in this paper provide
a deeper insight into the microscopic mechanism of cavity driving
and the entanglement phenomena in conventional cavity optomechanical
systems.
\begin{acknowledgments}
This work was supported by the National Natural Science Foundation
of China (No. 12365005, No.12265024). M. A. Bake acknowledges support from the Natural Science Foundation of Xinjiang China (Grant No. 2022D01C421).
\end{acknowledgments}

\appendix

\section{\label{sec:A1}Relevant expressions}

\begin{widetext}

In this paper, we have derived explicit expressions for the relevant
quantities of our system. However, due to the expressions of $A(\omega)$,
$B(\omega)$, $C(\omega)$, $D(\omega)$, and $F(\omega)$ are too
cumbersome to include in the main text. Therefore, we have provided
them in this Appendix for reference. After calculation we get the
$\delta c(\omega)$ as
\begin{equation}
\delta c(\omega)=A^{\prime}(\omega)\delta c_{in}(\omega)+B^{\prime}(\omega)\delta c_{in}^{\dagger}(-\omega)+C^{\prime}(\omega)\delta B_{in}(\omega)+D^{\prime}(\omega)\delta B_{in}^{\dagger}(-\omega)+F^{\prime}(\omega)\xi(\omega),
\end{equation}
 where
\begin{equation}
A(\omega)=\sqrt{2\kappa}A^{\prime}(\omega)-1,\quad B(\omega)=\sqrt{2\kappa}B^{\prime}(\omega)
\end{equation}

\begin{equation}
C(\omega)=\sqrt{2\kappa}C^{\prime}(\omega),\quad D(\omega)=\sqrt{2\kappa}D^{\prime}(\omega),\quad F(\omega)=\sqrt{2\kappa}F^{\prime}(\omega).
\end{equation}
 The expressions of $A^{\prime}(\omega)$, $B^{\prime}(\omega)$,
$C^{\prime}(\omega)$, $D^{\prime}(\omega)$, and $F^{\prime}(\omega)$
are listed below:
\begin{align}
A^{\prime}(\omega) & =\frac{\sqrt{2\kappa}}{d(\omega)}[iG_{0}^{2}\omega_{m}\left|c_{s}\right|{}^{2}\left(\left|G_{1}\right|{}^{2}-\nu_{1}\nu_{2}\right)-\left(\omega^{2}+i\omega\gamma_{m}\right)\left(\nu_{2}\left|G_{3}\right|{}^{2}-\nu_{2}\mu_{2}\nu_{1}+\mu_{2}\left|G_{1}\right|{}^{2}\right)\nonumber \\
 & +\omega_{m}^{2}\left(\nu_{2}\left|G_{3}\right|{}^{2}-\nu_{2}\mu_{2}\nu_{1}+\mu_{2}\left|G_{1}\right|{}^{2}\right)+\left(G_{1}G_{3}^{*}G_{2}^{*}+G_{3}G_{1}^{*}G_{2}^{*}+i\nu_{1}G_{2}^{*2}\right)\left(\omega\gamma_{m}-i\omega^{2}+i\omega_{m}^{2}\right)],
\end{align}
\begin{equation}
B^{\prime}(\omega)=\frac{\sqrt{2\kappa}i}{d(\omega)}[G_{0}^{2}c_{s}^{2}\omega_{m}\left(\left|G_{1}\right|{}^{2}-\nu_{1}\nu_{2}\right)+\left(i\omega\gamma_{m}-\omega_{m}^{2}+\omega^{2}\right)\left(G_{1}\left|G_{2}\right|^{2}+iG_{3}G_{2}^{*}\nu_{1}+G_{3}^{2}G_{1}^{*}-iG_{2}G_{3}\nu_{2}\right)],
\end{equation}

\begin{align}
C^{\prime}(\omega) & =\frac{\sqrt{2\gamma_{a}}i}{d(\omega)}[-G_{0}^{2}\omega_{m}\left(G_{1}^{*}G_{3}\left|c_{s}\right|{}^{2}+c_{s}^{2}G_{1}^{*}G_{2}^{*}-i\nu_{2}\left(G_{2}\left|c_{s}\right|{}^{2}+c_{s}^{2}G_{3}^{*}\right)\right)\nonumber \\
 & +(\omega^{2}+i\omega\gamma_{m}-\omega_{m}^{2})\left(\left|G_{3}\right|{}^{2}G_{2}^{*}-\left|G_{2}\right|{}^{2}G_{2}^{*}-\mu_{2}\nu_{2}G_{2}-iG_{3}\mu_{2}G_{1}^{*}\right)],
\end{align}

\begin{align}
D^{\prime}(\omega) & =\frac{\sqrt{2\gamma_{a}}}{d(\omega)}[G_{2}^{*}G_{0}^{2}c_{s}^{2}\nu_{1}\omega_{m}+\left(i\omega\gamma_{m}-\omega_{m}^{2}+\omega^{2}\right)\left(i\left|G_{2}\right|{}^{2}G_{3}+G_{1}G_{2}\mu_{2}-i\left|G_{3}\right|{}^{2}G_{3}-iG_{3}\nu_{1}\mu_{2}\right)\nonumber \\
 & -iG_{3}^{*}G_{1}G_{0}^{2}c_{s}^{2}\omega_{m}+G_{0}^{2}\omega_{m}\left|c_{s}\right|{}^{2}\left(G_{3}\nu_{1}-iG_{1}G_{2}\right)],
\end{align}

\begin{align}
F^{\prime}(\omega) & =\frac{iG_{0}\omega_{m}}{d(\omega)}[c_{s}\left(\nu_{2}(\left|G_{3}\right|{}^{2}-\mu_{2}\nu_{1})+\mu_{2}\left|G_{1}\right|{}^{2}-\nu_{1}G_{2}^{*2}\right)+i\left|G_{2}\right|{}^{2}G_{1}c_{s}^{*}\nonumber \\
 & +iG_{2}^{*}G_{1}G_{3}^{*}c_{s}+G_{2}^{*}G_{3}(iG_{1}^{*}c_{s}-\nu_{1}c_{s}^{*})+G_{3}c_{s}^{*}(G_{2}\nu_{2}+iG_{3}G_{1}^{*})],
\end{align}

\begin{align}
d(\omega) & =-2i\omega\left|G_{2}G_{3}\right|{}^{2}\gamma_{m}-\mu_{1}\omega(i\omega+\gamma_{m})\left(G_{1}G_{2}^{*}G_{3}^{*}-G_{3}G_{1}^{*}G_{2}^{*}\right)+\mu_{2}\omega(i\omega-\gamma_{m})\left(G_{1}G_{2}G_{3}^{*}+G_{2}G_{3}G_{1}^{*}\right)\nonumber \\
 & -\omega(\omega+i\gamma_{m})\left(\mu_{1}\mu_{2}\left|G_{1}\right|{}^{2}-\mu_{1}\nu_{1}G_{2}^{*2}-G_{2}^{2}\mu_{2}\nu_{2}-\mu_{1}\mu_{2}\nu_{1}\nu_{2}\right)\nonumber \\
 & +iG_{0}^{2}\left(c_{s}^{2}G_{3}^{*}\left(\nu_{1}G_{2}^{*}-G_{2}\nu_{2}\right)-iG_{1}c_{s}^{2}G_{3}^{*2}+G_{3}c_{s}^{2}\left(\nu_{1}G_{2}^{*}-iG_{3}G_{1}^{*}-G_{2}\nu_{2}\right)\right.\nonumber \\
 & +\left|c_{s}\right|{}^{2}\left(\left(\mu_{1}-\mu_{2}\right)\left(\left|G_{1}\right|{}^{2}-\nu_{1}\nu_{2}\right)+\nu_{1}G_{2}^{*2}-2i\left(G_{1}G_{3}^{*}+G_{3}G_{1}^{*}\right)\Re\left(G_{2}\right)-G_{2}^{2}\nu_{2}\right)\omega_{m}\nonumber \\
 & +(2\left|G_{2}G_{3}\right|{}^{2}-\mu_{1}\nu_{1}G_{2}^{*2}+i\mu_{1}G_{2}^{*}\left(G_{1}G_{3}^{*}+G_{3}G_{1}^{*}\right)\nonumber \\
 & +\mu_{2}\left(\mu_{1}\left|G_{1}\right|{}^{2}-iG_{2}\left(G_{1}G_{3}^{*}+G_{3}G_{1}^{*}\right)-\nu_{2}\left(G_{2}^{2}+\mu_{1}\nu_{1}\right)\right)\omega_{m}^{2}\nonumber \\
 & +\left|G_{2}\right|{}^{2}\left(G_{0}^{2}\omega_{m}\left(G_{1}c_{s}^{*2}+c_{s}^{2}G_{1}^{*}\right)-2\omega^{2}\left|G_{3}\right|{}^{2}\right)+(\left|G_{2}\right|{}^{4}+\left|G_{3}\right|{}^{4})\left(i\omega\gamma_{m}-\omega_{m}^{2}+\omega^{2}\right)\nonumber \\
 & +\left|G_{3}\right|{}^{2}(i\left(\nu_{1}-\nu_{2}\right)G_{0}^{2}\omega_{m}\left|c_{s}\right|{}^{2}-\left(\mu_{2}\nu_{1}+\mu_{1}\nu_{2}\right)\left(\omega^{2}+i\gamma_{m}\omega\right)+\left(\mu_{2}\nu_{1}+\mu_{1}\nu_{2}\right)\omega_{m}^{2}).
\end{align}

\end{widetext}

\bibliographystyle{apsrev4-1}
\bibliography{Reference}

\begin{thebibliography}{56}%
\makeatletter
\providecommand \@ifxundefined [1]{%
 \@ifx{#1\undefined}
}%
\providecommand \@ifnum [1]{%
 \ifnum #1\expandafter \@firstoftwo
 \else \expandafter \@secondoftwo
 \fi
}%
\providecommand \@ifx [1]{%
 \ifx #1\expandafter \@firstoftwo
 \else \expandafter \@secondoftwo
 \fi
}%
\providecommand \natexlab [1]{#1}%
\providecommand \enquote  [1]{``#1''}%
\providecommand \bibnamefont  [1]{#1}%
\providecommand \bibfnamefont [1]{#1}%
\providecommand \citenamefont [1]{#1}%
\providecommand \href@noop [0]{\@secondoftwo}%
\providecommand \href [0]{\begingroup \@sanitize@url \@href}%
\providecommand \@href[1]{\@@startlink{#1}\@@href}%
\providecommand \@@href[1]{\endgroup#1\@@endlink}%
\providecommand \@sanitize@url [0]{\catcode `\\12\catcode `\$12\catcode
  `\&12\catcode `\#12\catcode `\^12\catcode `\_12\catcode `\%12\relax}%
\providecommand \@@startlink[1]{}%
\providecommand \@@endlink[0]{}%
\providecommand \url  [0]{\begingroup\@sanitize@url \@url }%
\providecommand \@url [1]{\endgroup\@href {#1}{\urlprefix }}%
\providecommand \urlprefix  [0]{URL }%
\providecommand \Eprint [0]{\href }%
\providecommand \doibase [0]{http://dx.doi.org/}%
\providecommand \selectlanguage [0]{\@gobble}%
\providecommand \bibinfo  [0]{\@secondoftwo}%
\providecommand \bibfield  [0]{\@secondoftwo}%
\providecommand \translation [1]{[#1]}%
\providecommand \BibitemOpen [0]{}%
\providecommand \bibitemStop [0]{}%
\providecommand \bibitemNoStop [0]{.\EOS\space}%
\providecommand \EOS [0]{\spacefactor3000\relax}%
\providecommand \BibitemShut  [1]{\csname bibitem#1\endcsname}%
\let\auto@bib@innerbib\@empty
\bibitem [{\citenamefont {Amico}\ \emph {et~al.}(2008)\citenamefont {Amico},
  \citenamefont {Fazio}, \citenamefont {Osterloh},\ and\ \citenamefont
  {Vedral}}]{RevModPhys.80.517}%
  \BibitemOpen
  \bibfield  {author} {\bibinfo {author} {\bibfnamefont {L.}~\bibnamefont
  {Amico}}, \bibinfo {author} {\bibfnamefont {R.}~\bibnamefont {Fazio}},
  \bibinfo {author} {\bibfnamefont {A.}~\bibnamefont {Osterloh}}, \ and\
  \bibinfo {author} {\bibfnamefont {V.}~\bibnamefont {Vedral}},\ }Entanglement
  in many-body systems,\ \href {\doibase 10.1103/RevModPhys.80.517} {\bibfield
  {journal} {\bibinfo  {journal} {Rev. Mod. Phys.}\ }\textbf {\bibinfo {volume}
  {80}},\ \bibinfo {pages} {517} (\bibinfo {year} {2008})}\BibitemShut
  {NoStop}%
\bibitem [{\citenamefont {Horodecki}\ \emph {et~al.}(2009)\citenamefont
  {Horodecki}, \citenamefont {Horodecki}, \citenamefont {Horodecki},\ and\
  \citenamefont {Horodecki}}]{RevModPhys.81.865}%
  \BibitemOpen
  \bibfield  {author} {\bibinfo {author} {\bibfnamefont {R.}~\bibnamefont
  {Horodecki}}, \bibinfo {author} {\bibfnamefont {P.}~\bibnamefont
  {Horodecki}}, \bibinfo {author} {\bibfnamefont {M.}~\bibnamefont
  {Horodecki}}, \ and\ \bibinfo {author} {\bibfnamefont {K.}~\bibnamefont
  {Horodecki}},\ }Quantum entanglement,\ \href {\doibase
  10.1103/RevModPhys.81.865} {\bibfield  {journal} {\bibinfo  {journal} {Rev.
  Mod. Phys.}\ }\textbf {\bibinfo {volume} {81}},\ \bibinfo {pages} {865}
  (\bibinfo {year} {2009})}\BibitemShut {NoStop}%
\bibitem [{\citenamefont {Nielsen}\ and\ \citenamefont
  {Chuang}(2010)}]{Nielsen_Chuang_2010}%
  \BibitemOpen
  \bibfield  {author} {\bibinfo {author} {\bibfnamefont {M.~A.}\ \bibnamefont
  {Nielsen}}\ and\ \bibinfo {author} {\bibfnamefont {I.~L.}\ \bibnamefont
  {Chuang}},\ }\href@noop {} {\emph {\bibinfo {title} {Quantum Computation and
  Quantum Information: 10th Anniversary Edition}}}\ (\bibinfo  {publisher}
  {Cambridge University Press},\ \bibinfo {year} {2010})\BibitemShut {NoStop}%
\bibitem [{\citenamefont {Pan}\ \emph {et~al.}(2012)\citenamefont {Pan},
  \citenamefont {Chen}, \citenamefont {Lu}, \citenamefont {Weinfurter},
  \citenamefont {Zeilinger},\ and\ \citenamefont {\ifmmode~\dot{Z}\else
  \.{Z}\fi{}ukowski}}]{RevModPhys.84.777}%
  \BibitemOpen
  \bibfield  {author} {\bibinfo {author} {\bibfnamefont {J.-W.}\ \bibnamefont
  {Pan}}, \bibinfo {author} {\bibfnamefont {Z.-B.}\ \bibnamefont {Chen}},
  \bibinfo {author} {\bibfnamefont {C.-Y.}\ \bibnamefont {Lu}}, \bibinfo
  {author} {\bibfnamefont {H.}~\bibnamefont {Weinfurter}}, \bibinfo {author}
  {\bibfnamefont {A.}~\bibnamefont {Zeilinger}}, \ and\ \bibinfo {author}
  {\bibfnamefont {M.}~\bibnamefont {\ifmmode~\dot{Z}\else \.{Z}\fi{}ukowski}},\
  }Multiphoton entanglement and interferometry,\ \href {\doibase
  10.1103/RevModPhys.84.777} {\bibfield  {journal} {\bibinfo  {journal} {Rev.
  Mod. Phys.}\ }\textbf {\bibinfo {volume} {84}},\ \bibinfo {pages} {777}
  (\bibinfo {year} {2012})}\BibitemShut {NoStop}%
\bibitem [{\citenamefont {Pezz\`e}\ \emph {et~al.}(2018)\citenamefont
  {Pezz\`e}, \citenamefont {Smerzi}, \citenamefont {Oberthaler}, \citenamefont
  {Schmied},\ and\ \citenamefont {Treutlein}}]{RevModPhys.90.035005}%
  \BibitemOpen
  \bibfield  {author} {\bibinfo {author} {\bibfnamefont {L.}~\bibnamefont
  {Pezz\`e}}, \bibinfo {author} {\bibfnamefont {A.}~\bibnamefont {Smerzi}},
  \bibinfo {author} {\bibfnamefont {M.~K.}\ \bibnamefont {Oberthaler}},
  \bibinfo {author} {\bibfnamefont {R.}~\bibnamefont {Schmied}}, \ and\
  \bibinfo {author} {\bibfnamefont {P.}~\bibnamefont {Treutlein}},\ }Quantum
  metrology with nonclassical states of atomic ensembles,\ \href {\doibase
  10.1103/RevModPhys.90.035005} {\bibfield  {journal} {\bibinfo  {journal}
  {Rev. Mod. Phys.}\ }\textbf {\bibinfo {volume} {90}},\ \bibinfo {pages}
  {035005} (\bibinfo {year} {2018})}\BibitemShut {NoStop}%
\bibitem [{\citenamefont {Monroe}\ \emph {et~al.}(1995)\citenamefont {Monroe},
  \citenamefont {Meekhof}, \citenamefont {King}, \citenamefont {Itano},\ and\
  \citenamefont {Wineland}}]{PhysRevLett.75.4714}%
  \BibitemOpen
  \bibfield  {author} {\bibinfo {author} {\bibfnamefont {C.}~\bibnamefont
  {Monroe}}, \bibinfo {author} {\bibfnamefont {D.~M.}\ \bibnamefont {Meekhof}},
  \bibinfo {author} {\bibfnamefont {B.~E.}\ \bibnamefont {King}}, \bibinfo
  {author} {\bibfnamefont {W.~M.}\ \bibnamefont {Itano}}, \ and\ \bibinfo
  {author} {\bibfnamefont {D.~J.}\ \bibnamefont {Wineland}},\ }Demonstration of
  a Fundamental Quantum Logic Gate,\ \href {\doibase
  10.1103/PhysRevLett.75.4714} {\bibfield  {journal} {\bibinfo  {journal}
  {Phys. Rev. Lett.}\ }\textbf {\bibinfo {volume} {75}},\ \bibinfo {pages}
  {4714} (\bibinfo {year} {1995})}\BibitemShut {NoStop}%
\bibitem [{\citenamefont {Leibfried}\ \emph {et~al.}(2003)\citenamefont
  {Leibfried}, \citenamefont {Blatt}, \citenamefont {Monroe},\ and\
  \citenamefont {Wineland}}]{RevModPhys.75.281}%
  \BibitemOpen
  \bibfield  {author} {\bibinfo {author} {\bibfnamefont {D.}~\bibnamefont
  {Leibfried}}, \bibinfo {author} {\bibfnamefont {R.}~\bibnamefont {Blatt}},
  \bibinfo {author} {\bibfnamefont {C.}~\bibnamefont {Monroe}}, \ and\ \bibinfo
  {author} {\bibfnamefont {D.}~\bibnamefont {Wineland}},\ }Quantum dynamics of
  single trapped ions,\ \href {\doibase 10.1103/RevModPhys.75.281} {\bibfield
  {journal} {\bibinfo  {journal} {Rev. Mod. Phys.}\ }\textbf {\bibinfo {volume}
  {75}},\ \bibinfo {pages} {281} (\bibinfo {year} {2003})}\BibitemShut
  {NoStop}%
\bibitem [{\citenamefont {Brownnutt}\ \emph {et~al.}(2015)\citenamefont
  {Brownnutt}, \citenamefont {Kumph}, \citenamefont {Rabl},\ and\ \citenamefont
  {Blatt}}]{RevModPhys.87.1419}%
  \BibitemOpen
  \bibfield  {author} {\bibinfo {author} {\bibfnamefont {M.}~\bibnamefont
  {Brownnutt}}, \bibinfo {author} {\bibfnamefont {M.}~\bibnamefont {Kumph}},
  \bibinfo {author} {\bibfnamefont {P.}~\bibnamefont {Rabl}}, \ and\ \bibinfo
  {author} {\bibfnamefont {R.}~\bibnamefont {Blatt}},\ }Ion-trap measurements
  of electric-field noise near surfaces,\ \href {\doibase
  10.1103/RevModPhys.87.1419} {\bibfield  {journal} {\bibinfo  {journal} {Rev.
  Mod. Phys.}\ }\textbf {\bibinfo {volume} {87}},\ \bibinfo {pages} {1419}
  (\bibinfo {year} {2015})}\BibitemShut {NoStop}%
\bibitem [{\citenamefont {Reiserer}\ and\ \citenamefont
  {Rempe}(2015)}]{RevModPhys.87.1379}%
  \BibitemOpen
  \bibfield  {author} {\bibinfo {author} {\bibfnamefont {A.}~\bibnamefont
  {Reiserer}}\ and\ \bibinfo {author} {\bibfnamefont {G.}~\bibnamefont
  {Rempe}},\ }Cavity-based quantum networks with single atoms and optical
  photons,\ \href {\doibase 10.1103/RevModPhys.87.1379} {\bibfield  {journal}
  {\bibinfo  {journal} {Rev. Mod. Phys.}\ }\textbf {\bibinfo {volume} {87}},\
  \bibinfo {pages} {1379} (\bibinfo {year} {2015})}\BibitemShut {NoStop}%
\bibitem [{\citenamefont {Blais}\ \emph {et~al.}(2021)\citenamefont {Blais},
  \citenamefont {Grimsmo}, \citenamefont {Girvin},\ and\ \citenamefont
  {Wallraff}}]{RevModPhys.93.025005}%
  \BibitemOpen
  \bibfield  {author} {\bibinfo {author} {\bibfnamefont {A.}~\bibnamefont
  {Blais}}, \bibinfo {author} {\bibfnamefont {A.~L.}\ \bibnamefont {Grimsmo}},
  \bibinfo {author} {\bibfnamefont {S.~M.}\ \bibnamefont {Girvin}}, \ and\
  \bibinfo {author} {\bibfnamefont {A.}~\bibnamefont {Wallraff}},\ }Circuit
  quantum electrodynamics,\ \href {\doibase 10.1103/RevModPhys.93.025005}
  {\bibfield  {journal} {\bibinfo  {journal} {Rev. Mod. Phys.}\ }\textbf
  {\bibinfo {volume} {93}},\ \bibinfo {pages} {025005} (\bibinfo {year}
  {2021})}\BibitemShut {NoStop}%
\bibitem [{\citenamefont {Aspelmeyer}\ \emph
  {et~al.}(2014{\natexlab{a}})\citenamefont {Aspelmeyer}, \citenamefont
  {Kippenberg},\ and\ \citenamefont {Marquardt}}]{RevModPhys.86.1391}%
  \BibitemOpen
  \bibfield  {author} {\bibinfo {author} {\bibfnamefont {M.}~\bibnamefont
  {Aspelmeyer}}, \bibinfo {author} {\bibfnamefont {T.~J.}\ \bibnamefont
  {Kippenberg}}, \ and\ \bibinfo {author} {\bibfnamefont {F.}~\bibnamefont
  {Marquardt}},\ }Cavity optomechanics,\ \href {\doibase
  10.1103/RevModPhys.86.1391} {\bibfield  {journal} {\bibinfo  {journal} {Rev.
  Mod. Phys.}\ }\textbf {\bibinfo {volume} {86}},\ \bibinfo {pages} {1391}
  (\bibinfo {year} {2014}{\natexlab{a}})}\BibitemShut {NoStop}%
\bibitem [{\citenamefont {M.~Aspelmeyer}\ and\ \citenamefont
  {Marquardt}(2014)}]{b1}%
  \BibitemOpen
  \bibfield  {author} {\bibinfo {author} {\bibfnamefont {T.~J.~K.}\
  \bibnamefont {M.~Aspelmeyer}}\ and\ \bibinfo {author} {\bibfnamefont
  {F.}~\bibnamefont {Marquardt}},\ }\href {\doibase
  doi:10.1007/978-3-642-55312-7} {} {\emph {\bibinfo {title} {in Cavity
  Optomechanics: Nano- and Micromechanical
  Resonators Interacting with
  Light}}}\ (\bibinfo  {publisher} {Springer},\ \bibinfo {address}
  {Heidelberg},\ \bibinfo {year} {2014})\BibitemShut {NoStop}%
\bibitem [{\citenamefont {Bowen}\ and\ \citenamefont {Milburn}(2016)}]{b2}%
  \BibitemOpen
  \bibfield  {author} {\bibinfo {author} {\bibfnamefont {W.~B.}\ \bibnamefont
  {Bowen}}\ and\ \bibinfo {author} {\bibfnamefont {G.~J.}\ \bibnamefont
  {Milburn}},\ }\href@noop {} {\emph {\bibinfo {title} {Quantum
  Optomechanics}}}\ (\bibinfo  {publisher} {CRC Press},\ \bibinfo {address}
  {Boca Raton, FL},\ \bibinfo {year} {2016})\BibitemShut {NoStop}%
\bibitem [{\citenamefont {Zuo}\ \emph {et~al.}(2024)\citenamefont {Zuo},
  \citenamefont {Fan}, \citenamefont {Qian}, \citenamefont {Ding},
  \citenamefont {Tan}, \citenamefont {Xiong},\ and\ \citenamefont
  {Li}}]{Zuo_2024}%
  \BibitemOpen
  \bibfield  {author} {\bibinfo {author} {\bibfnamefont {X.}~\bibnamefont
  {Zuo}}, \bibinfo {author} {\bibfnamefont {Z.-Y.}\ \bibnamefont {Fan}},
  \bibinfo {author} {\bibfnamefont {H.}~\bibnamefont {Qian}}, \bibinfo {author}
  {\bibfnamefont {M.-S.}\ \bibnamefont {Ding}}, \bibinfo {author}
  {\bibfnamefont {H.}~\bibnamefont {Tan}}, \bibinfo {author} {\bibfnamefont
  {H.}~\bibnamefont {Xiong}}, \ and\ \bibinfo {author} {\bibfnamefont
  {J.}~\bibnamefont {Li}},\ }Cavity magnomechanics: from classical to quantum,\
  \href {\doibase 10.1088/1367-2630/ad327c} {\bibfield  {journal} {\bibinfo
  {journal} {New J. Phys.}\ }\textbf {\bibinfo {volume} {26}},\ \bibinfo
  {pages} {031201} (\bibinfo {year} {2024})}\BibitemShut {NoStop}%
\bibitem [{\citenamefont {Zurek}(2003)}]{RevModPhys.75.715}%
  \BibitemOpen
  \bibfield  {author} {\bibinfo {author} {\bibfnamefont {W.~H.}\ \bibnamefont
  {Zurek}},\ }Decoherence, einselection, and the quantum origins of the
  classical,\ \href {\doibase 10.1103/RevModPhys.75.715} {\bibfield  {journal}
  {\bibinfo  {journal} {Rev. Mod. Phys.}\ }\textbf {\bibinfo {volume} {75}},\
  \bibinfo {pages} {715} (\bibinfo {year} {2003})}\BibitemShut {NoStop}%
\bibitem [{\citenamefont {Brennecke}\ \emph {et~al.}(2008)\citenamefont
  {Brennecke}, \citenamefont {Ritter}, \citenamefont {Donner},\ and\
  \citenamefont {Esslinger}}]{A23}%
  \BibitemOpen
  \bibfield  {author} {\bibinfo {author} {\bibfnamefont {F.}~\bibnamefont
  {Brennecke}}, \bibinfo {author} {\bibfnamefont {S.}~\bibnamefont {Ritter}},
  \bibinfo {author} {\bibfnamefont {T.}~\bibnamefont {Donner}}, \ and\ \bibinfo
  {author} {\bibfnamefont {T.}~\bibnamefont {Esslinger}},\ }Cavity
  Optomechanics with a Bose-Einstein Condensate,\ \href {\doibase
  10.1126/science.1163218} {\bibfield  {journal} {\bibinfo  {journal}
  {Science}\ }\textbf {\bibinfo {volume} {322}},\ \bibinfo {pages} {235}
  (\bibinfo {year} {2008})}\BibitemShut {NoStop}%
\bibitem [{\citenamefont {Modi}\ \emph {et~al.}(2012)\citenamefont {Modi},
  \citenamefont {Brodutch}, \citenamefont {Cable}, \citenamefont {Paterek},\
  and\ \citenamefont {Vedral}}]{RevModPhys.84.1655}%
  \BibitemOpen
  \bibfield  {author} {\bibinfo {author} {\bibfnamefont {K.}~\bibnamefont
  {Modi}}, \bibinfo {author} {\bibfnamefont {A.}~\bibnamefont {Brodutch}},
  \bibinfo {author} {\bibfnamefont {H.}~\bibnamefont {Cable}}, \bibinfo
  {author} {\bibfnamefont {T.}~\bibnamefont {Paterek}}, \ and\ \bibinfo
  {author} {\bibfnamefont {V.}~\bibnamefont {Vedral}},\ }The classical-quantum
  boundary for correlations: Discord and related measures,\ \href {\doibase
  10.1103/RevModPhys.84.1655} {\bibfield  {journal} {\bibinfo  {journal} {Rev.
  Mod. Phys.}\ }\textbf {\bibinfo {volume} {84}},\ \bibinfo {pages} {1655}
  (\bibinfo {year} {2012})}\BibitemShut {NoStop}%
\bibitem [{\citenamefont {Haroche}(2013)}]{RevModPhys.85.1083}%
  \BibitemOpen
  \bibfield  {author} {\bibinfo {author} {\bibfnamefont {S.}~\bibnamefont
  {Haroche}},\ }Nobel Lecture: Controlling photons in a box and exploring the
  quantum to classical boundary,\ \href {\doibase 10.1103/RevModPhys.85.1083}
  {\bibfield  {journal} {\bibinfo  {journal} {Rev. Mod. Phys.}\ }\textbf
  {\bibinfo {volume} {85}},\ \bibinfo {pages} {1083} (\bibinfo {year}
  {2013})}\BibitemShut {NoStop}%
\bibitem [{\citenamefont {Chen}\ \emph
  {et~al.}(2017{\natexlab{a}})\citenamefont {Chen}, \citenamefont {Nie},
  \citenamefont {Li}, \citenamefont {Zeng}, \citenamefont {Liao},\ and\
  \citenamefont {Xiao}}]{CHEN201797}%
  \BibitemOpen
  \bibfield  {author} {\bibinfo {author} {\bibfnamefont {A.}~\bibnamefont
  {Chen}}, \bibinfo {author} {\bibfnamefont {W.}~\bibnamefont {Nie}}, \bibinfo
  {author} {\bibfnamefont {L.}~\bibnamefont {Li}}, \bibinfo {author}
  {\bibfnamefont {W.}~\bibnamefont {Zeng}}, \bibinfo {author} {\bibfnamefont
  {Q.}~\bibnamefont {Liao}}, \ and\ \bibinfo {author} {\bibfnamefont
  {X.}~\bibnamefont {Xiao}},\ }Steady-state entanglement in levitated
  optomechanical systems coupled to a higher order excited atomic ensemble,\
  \href {\doibase https://doi.org/10.1016/j.optcom.2017.06.103} {\bibfield
  {journal} {\bibinfo  {journal} {Opt. Commun.}\ }\textbf {\bibinfo {volume}
  {403}},\ \bibinfo {pages} {97} (\bibinfo {year}
  {2017}{\natexlab{a}})}\BibitemShut {NoStop}%
\bibitem [{\citenamefont {Liao}\ \emph {et~al.}(2018)\citenamefont {Liao},
  \citenamefont {Xie}, \citenamefont {Shang}, \citenamefont {Chen},\ and\
  \citenamefont {Lin}}]{Liao18}%
  \BibitemOpen
  \bibfield  {author} {\bibinfo {author} {\bibfnamefont {C.-G.}\ \bibnamefont
  {Liao}}, \bibinfo {author} {\bibfnamefont {H.}~\bibnamefont {Xie}}, \bibinfo
  {author} {\bibfnamefont {X.}~\bibnamefont {Shang}}, \bibinfo {author}
  {\bibfnamefont {Z.-H.}\ \bibnamefont {Chen}}, \ and\ \bibinfo {author}
  {\bibfnamefont {X.-M.}\ \bibnamefont {Lin}},\ }Enhancement of steady-state
  bosonic squeezing and entanglement in a dissipative optomechanical system,\
  \href {\doibase 10.1364/OE.26.013783} {\bibfield  {journal} {\bibinfo
  {journal} {Opt. Express}\ }\textbf {\bibinfo {volume} {26}},\ \bibinfo
  {pages} {13783} (\bibinfo {year} {2018})}\BibitemShut {NoStop}%
\bibitem [{\citenamefont {Di~Stefano}\ \emph {et~al.}(2019)\citenamefont
  {Di~Stefano}, \citenamefont {Settineri}, \citenamefont {Macr{\i}},
  \citenamefont {Ridolfo}, \citenamefont {Stassi}, \citenamefont {Kockum},
  \citenamefont {Savasta},\ and\ \citenamefont
  {Nori}}]{PhysRevLett.122.030402}%
  \BibitemOpen
  \bibfield  {author} {\bibinfo {author} {\bibfnamefont {O.}~\bibnamefont
  {Di~Stefano}}, \bibinfo {author} {\bibfnamefont {A.}~\bibnamefont
  {Settineri}}, \bibinfo {author} {\bibfnamefont {V.}~\bibnamefont {Macr{\i}}},
  \bibinfo {author} {\bibfnamefont {A.}~\bibnamefont {Ridolfo}}, \bibinfo
  {author} {\bibfnamefont {R.}~\bibnamefont {Stassi}}, \bibinfo {author}
  {\bibfnamefont {A.~F.}\ \bibnamefont {Kockum}}, \bibinfo {author}
  {\bibfnamefont {S.}~\bibnamefont {Savasta}}, \ and\ \bibinfo {author}
  {\bibfnamefont {F.}~\bibnamefont {Nori}},\ }Interaction of Mechanical
  Oscillators Mediated by the Exchange of Virtual Photon Pairs,\ \href
  {\doibase 10.1103/PhysRevLett.122.030402} {\bibfield  {journal} {\bibinfo
  {journal} {Phys. Rev. Lett.}\ }\textbf {\bibinfo {volume} {122}},\ \bibinfo
  {pages} {030402} (\bibinfo {year} {2019})}\BibitemShut {NoStop}%
\bibitem [{\citenamefont {Eftekhari}\ \emph {et~al.}(2022)\citenamefont
  {Eftekhari}, \citenamefont {Tavassoly},\ and\ \citenamefont
  {Behjat}}]{EFTEKHARI2022127176}%
  \BibitemOpen
  \bibfield  {author} {\bibinfo {author} {\bibfnamefont {F.}~\bibnamefont
  {Eftekhari}}, \bibinfo {author} {\bibfnamefont {M.}~\bibnamefont
  {Tavassoly}}, \ and\ \bibinfo {author} {\bibfnamefont {A.}~\bibnamefont
  {Behjat}},\ }Nonlinear interaction of a three-level atom with a two-mode
  field in an optomechanical cavity: Field and mechanical mode dissipations,\
  \href {\doibase https://doi.org/10.1016/j.physa.2022.127176} {\bibfield
  {journal} {\bibinfo  {journal} {Phys. A: Stat. Mech. Appl.}\ }\textbf
  {\bibinfo {volume} {596}},\ \bibinfo {pages} {127176} (\bibinfo {year}
  {2022})}\BibitemShut {NoStop}%
\bibitem [{\citenamefont {Zhang}\ \emph {et~al.}(2022)\citenamefont {Zhang},
  \citenamefont {Wang}, \citenamefont {Han}, \citenamefont {Zhang},\ and\
  \citenamefont {Wang}}]{ZHANG2022127824}%
  \BibitemOpen
  \bibfield  {author} {\bibinfo {author} {\bibfnamefont {W.}~\bibnamefont
  {Zhang}}, \bibinfo {author} {\bibfnamefont {T.}~\bibnamefont {Wang}},
  \bibinfo {author} {\bibfnamefont {X.}~\bibnamefont {Han}}, \bibinfo {author}
  {\bibfnamefont {S.}~\bibnamefont {Zhang}}, \ and\ \bibinfo {author}
  {\bibfnamefont {H.-F.}\ \bibnamefont {Wang}},\ }Mechanical squeezing induced
  by Duffing nonlinearity and two driving tones in an optomechanical system,\
  \href {\doibase https://doi.org/10.1016/j.physleta.2021.127824} {\bibfield
  {journal} {\bibinfo  {journal} {Phys. Lett. A}\ }\textbf {\bibinfo {volume}
  {424}},\ \bibinfo {pages} {127824} (\bibinfo {year} {2022})}\BibitemShut
  {NoStop}%
\bibitem [{\citenamefont {Jiao}\ \emph {et~al.}(2024)\citenamefont {Jiao},
  \citenamefont {Zuo}, \citenamefont {Wang}, \citenamefont {Lu}, \citenamefont
  {Liao}, \citenamefont {Kuang},\ and\ \citenamefont
  {Jing}}]{https://doi.org/10.1002/lpor.202301154}%
  \BibitemOpen
  \bibfield  {author} {\bibinfo {author} {\bibfnamefont {Y.-F.}\ \bibnamefont
  {Jiao}}, \bibinfo {author} {\bibfnamefont {Y.-L.}\ \bibnamefont {Zuo}},
  \bibinfo {author} {\bibfnamefont {Y.}~\bibnamefont {Wang}}, \bibinfo {author}
  {\bibfnamefont {W.}~\bibnamefont {Lu}}, \bibinfo {author} {\bibfnamefont
  {J.-Q.}\ \bibnamefont {Liao}}, \bibinfo {author} {\bibfnamefont {L.-M.}\
  \bibnamefont {Kuang}}, \ and\ \bibinfo {author} {\bibfnamefont
  {H.}~\bibnamefont {Jing}},\ }Tripartite Quantum Entanglement with Squeezed
  Optomechanics,\ \href {\doibase https://doi.org/10.1002/lpor.202301154}
  {\bibfield  {journal} {\bibinfo  {journal} {Laser Photonics Rev.}\ }\textbf
  {\bibinfo {volume} {18}},\ \bibinfo {pages} {2301154} (\bibinfo {year}
  {2024})}\BibitemShut {NoStop}%
\bibitem [{\citenamefont {Liao}\ and\ \citenamefont
  {Law}(2011)}]{PhysRevA.83.033820}%
  \BibitemOpen
  \bibfield  {author} {\bibinfo {author} {\bibfnamefont {J.-Q.}\ \bibnamefont
  {Liao}}\ and\ \bibinfo {author} {\bibfnamefont {C.~K.}\ \bibnamefont {Law}},\
  }Parametric generation of quadrature squeezing of mirrors in cavity
  optomechanics,\ \href {\doibase 10.1103/PhysRevA.83.033820} {\bibfield
  {journal} {\bibinfo  {journal} {Phys. Rev. A}\ }\textbf {\bibinfo {volume}
  {83}},\ \bibinfo {pages} {033820} (\bibinfo {year} {2011})}\BibitemShut
  {NoStop}%
\bibitem [{\citenamefont {Kronwald}\ \emph {et~al.}(2013)\citenamefont
  {Kronwald}, \citenamefont {Marquardt},\ and\ \citenamefont
  {Clerk}}]{PhysRevA.88.063833}%
  \BibitemOpen
  \bibfield  {author} {\bibinfo {author} {\bibfnamefont {A.}~\bibnamefont
  {Kronwald}}, \bibinfo {author} {\bibfnamefont {F.}~\bibnamefont {Marquardt}},
  \ and\ \bibinfo {author} {\bibfnamefont {A.~A.}\ \bibnamefont {Clerk}},\
  }Arbitrarily large steady-state bosonic squeezing via dissipation,\ \href
  {\doibase 10.1103/PhysRevA.88.063833} {\bibfield  {journal} {\bibinfo
  {journal} {Phys. Rev. A}\ }\textbf {\bibinfo {volume} {88}},\ \bibinfo
  {pages} {063833} (\bibinfo {year} {2013})}\BibitemShut {NoStop}%
\bibitem [{\citenamefont {J{\"a}hne}\ \emph {et~al.}(2009)\citenamefont
  {J{\"a}hne}, \citenamefont {Genes}, \citenamefont {Hammerer}, \citenamefont
  {Wallquist}, \citenamefont {Polzik},\ and\ \citenamefont
  {Zoller}}]{PhysRevA.79.063819}%
  \BibitemOpen
  \bibfield  {author} {\bibinfo {author} {\bibfnamefont {K.}~\bibnamefont
  {J{\"a}hne}}, \bibinfo {author} {\bibfnamefont {C.}~\bibnamefont {Genes}},
  \bibinfo {author} {\bibfnamefont {K.}~\bibnamefont {Hammerer}}, \bibinfo
  {author} {\bibfnamefont {M.}~\bibnamefont {Wallquist}}, \bibinfo {author}
  {\bibfnamefont {E.~S.}\ \bibnamefont {Polzik}}, \ and\ \bibinfo {author}
  {\bibfnamefont {P.}~\bibnamefont {Zoller}},\ }Cavity-assisted squeezing of a
  mechanical oscillator,\ \href {\doibase 10.1103/PhysRevA.79.063819}
  {\bibfield  {journal} {\bibinfo  {journal} {Phys. Rev. A}\ }\textbf {\bibinfo
  {volume} {79}},\ \bibinfo {pages} {063819} (\bibinfo {year}
  {2009})}\BibitemShut {NoStop}%
\bibitem [{\citenamefont {Huang}\ and\ \citenamefont
  {Agarwal}(2010)}]{PhysRevA.82.033811}%
  \BibitemOpen
  \bibfield  {author} {\bibinfo {author} {\bibfnamefont {S.}~\bibnamefont
  {Huang}}\ and\ \bibinfo {author} {\bibfnamefont {G.~S.}\ \bibnamefont
  {Agarwal}},\ }Reactive coupling can beat the motional quantum limit of
  nanowaveguides coupled to a microdisk resonator,\ \href {\doibase
  10.1103/PhysRevA.82.033811} {\bibfield  {journal} {\bibinfo  {journal} {Phys.
  Rev. A}\ }\textbf {\bibinfo {volume} {82}},\ \bibinfo {pages} {033811}
  (\bibinfo {year} {2010})}\BibitemShut {NoStop}%
\bibitem [{\citenamefont {Xiao}\ \emph {et~al.}(2014)\citenamefont {Xiao},
  \citenamefont {Yu},\ and\ \citenamefont {Zhang}}]{Xiao:14}%
  \BibitemOpen
  \bibfield  {author} {\bibinfo {author} {\bibfnamefont {Y.}~\bibnamefont
  {Xiao}}, \bibinfo {author} {\bibfnamefont {Y.-F.}\ \bibnamefont {Yu}}, \ and\
  \bibinfo {author} {\bibfnamefont {Z.-M.}\ \bibnamefont {Zhang}},\
  }Controllable optomechanically induced transparency and ponderomotive
  squeezing in an optomechanical system assisted by an atomic ensemble,\ \href
  {\doibase 10.1364/OE.22.017979} {\bibfield  {journal} {\bibinfo  {journal}
  {Opt. Express}\ }\textbf {\bibinfo {volume} {22}},\ \bibinfo {pages} {17979}
  (\bibinfo {year} {2014})}\BibitemShut {NoStop}%
\bibitem [{\citenamefont {Yu}\ and\ \citenamefont {Pan}(2024)}]{Yu_2024}%
  \BibitemOpen
  \bibfield  {author} {\bibinfo {author} {\bibfnamefont {G.}~\bibnamefont
  {Yu}}\ and\ \bibinfo {author} {\bibfnamefont {G.}~\bibnamefont {Pan}},\
  }Optomechanically induced transparency in a hybrid system containing
  two-level atomic ensemble and optical parametric amplifier,\ \href {\doibase
  10.1088/1555-6611/ad2beb} {\bibfield  {journal} {\bibinfo  {journal} {Laser
  Physics}\ }\textbf {\bibinfo {volume} {34}},\ \bibinfo {pages} {045203}
  (\bibinfo {year} {2024})}\BibitemShut {NoStop}%
\bibitem [{\citenamefont {Vitali}\ \emph {et~al.}(2007)\citenamefont {Vitali},
  \citenamefont {Gigan}, \citenamefont {Ferreira}, \citenamefont {B{\"o}hm},
  \citenamefont {Tombesi}, \citenamefont {Guerreiro}, \citenamefont {Vedral},
  \citenamefont {Zeilinger},\ and\ \citenamefont
  {Aspelmeyer}}]{PhysRevLett.98.030405}%
  \BibitemOpen
  \bibfield  {author} {\bibinfo {author} {\bibfnamefont {D.}~\bibnamefont
  {Vitali}}, \bibinfo {author} {\bibfnamefont {S.}~\bibnamefont {Gigan}},
  \bibinfo {author} {\bibfnamefont {A.}~\bibnamefont {Ferreira}}, \bibinfo
  {author} {\bibfnamefont {H.~R.}\ \bibnamefont {B{\"o}hm}}, \bibinfo {author}
  {\bibfnamefont {P.}~\bibnamefont {Tombesi}}, \bibinfo {author} {\bibfnamefont
  {A.}~\bibnamefont {Guerreiro}}, \bibinfo {author} {\bibfnamefont
  {V.}~\bibnamefont {Vedral}}, \bibinfo {author} {\bibfnamefont
  {A.}~\bibnamefont {Zeilinger}}, \ and\ \bibinfo {author} {\bibfnamefont
  {M.}~\bibnamefont {Aspelmeyer}},\ }Optomechanical Entanglement between a
  Movable Mirror and a Cavity Field,\ \href {\doibase
  10.1103/PhysRevLett.98.030405} {\bibfield  {journal} {\bibinfo  {journal}
  {Phys. Rev. Lett.}\ }\textbf {\bibinfo {volume} {98}},\ \bibinfo {pages}
  {030405} (\bibinfo {year} {2007})}\BibitemShut {NoStop}%
\bibitem [{\citenamefont {Stannigel}\ \emph {et~al.}(2012)\citenamefont
  {Stannigel}, \citenamefont {Komar}, \citenamefont {Habraken}, \citenamefont
  {Bennett}, \citenamefont {Lukin}, \citenamefont {Zoller},\ and\ \citenamefont
  {Rabl}}]{PhysRevLett.109.013603}%
  \BibitemOpen
  \bibfield  {author} {\bibinfo {author} {\bibfnamefont {K.}~\bibnamefont
  {Stannigel}}, \bibinfo {author} {\bibfnamefont {P.}~\bibnamefont {Komar}},
  \bibinfo {author} {\bibfnamefont {S.~J.~M.}\ \bibnamefont {Habraken}},
  \bibinfo {author} {\bibfnamefont {S.~D.}\ \bibnamefont {Bennett}}, \bibinfo
  {author} {\bibfnamefont {M.~D.}\ \bibnamefont {Lukin}}, \bibinfo {author}
  {\bibfnamefont {P.}~\bibnamefont {Zoller}}, \ and\ \bibinfo {author}
  {\bibfnamefont {P.}~\bibnamefont {Rabl}},\ }Optomechanical Quantum
  Information Processing with Photons and Phonons,\ \href {\doibase
  10.1103/PhysRevLett.109.013603} {\bibfield  {journal} {\bibinfo  {journal}
  {Phys. Rev. Lett.}\ }\textbf {\bibinfo {volume} {109}},\ \bibinfo {pages}
  {013603} (\bibinfo {year} {2012})}\BibitemShut {NoStop}%
\bibitem [{\citenamefont {Metcalfe}(2014)}]{4896}%
  \BibitemOpen
  \bibfield  {author} {\bibinfo {author} {\bibfnamefont {M.}~\bibnamefont
  {Metcalfe}},\ }Applications of cavity optomechanics,\ \href {\doibase
  10.1063/1.4896029} {\bibfield  {journal} {\bibinfo  {journal} {Appl. Phys.
  Rev.}\ }\textbf {\bibinfo {volume} {1}},\ \bibinfo {pages} {031105} (\bibinfo
  {year} {2014})}\BibitemShut {NoStop}%
\bibitem [{\citenamefont {Gupta}\ \emph {et~al.}(2007)\citenamefont {Gupta},
  \citenamefont {Moore}, \citenamefont {Murch},\ and\ \citenamefont
  {Stamper-Kurn}}]{PhysRevLett.99.213601}%
  \BibitemOpen
  \bibfield  {author} {\bibinfo {author} {\bibfnamefont {S.}~\bibnamefont
  {Gupta}}, \bibinfo {author} {\bibfnamefont {K.~L.}\ \bibnamefont {Moore}},
  \bibinfo {author} {\bibfnamefont {K.~W.}\ \bibnamefont {Murch}}, \ and\
  \bibinfo {author} {\bibfnamefont {D.~M.}\ \bibnamefont {Stamper-Kurn}},\
  }Cavity Nonlinear Optics at Low Photon Numbers from Collective Atomic
  Motion,\ \href {\doibase 10.1103/PhysRevLett.99.213601} {\bibfield  {journal}
  {\bibinfo  {journal} {Phys. Rev. Lett.}\ }\textbf {\bibinfo {volume} {99}},\
  \bibinfo {pages} {213601} (\bibinfo {year} {2007})}\BibitemShut {NoStop}%
\bibitem [{\citenamefont {Pirandola}\ \emph {et~al.}(2006)\citenamefont
  {Pirandola}, \citenamefont {Vitali}, \citenamefont {Tombesi},\ and\
  \citenamefont {Lloyd}}]{Pirandola_2006}%
  \BibitemOpen
  \bibfield  {author} {\bibinfo {author} {\bibfnamefont {S.}~\bibnamefont
  {Pirandola}}, \bibinfo {author} {\bibfnamefont {D.}~\bibnamefont {Vitali}},
  \bibinfo {author} {\bibfnamefont {P.}~\bibnamefont {Tombesi}}, \ and\
  \bibinfo {author} {\bibfnamefont {S.}~\bibnamefont {Lloyd}},\ }Macroscopic
  Entanglement by Entanglement Swapping,\ \href {\doibase
  10.1103/PhysRevLett.97.150403} {\bibfield  {journal} {\bibinfo  {journal}
  {Phys. Rev. Lett.}\ }\textbf {\bibinfo {volume} {97}},\ \bibinfo {pages}
  {150403} (\bibinfo {year} {2006})}\BibitemShut {NoStop}%
\bibitem [{\citenamefont {Aspelmeyer}\ \emph
  {et~al.}(2014{\natexlab{b}})\citenamefont {Aspelmeyer}, \citenamefont
  {Kippenberg},\ and\ \citenamefont {Marquardt}}]{Aspelmeyer2014}%
  \BibitemOpen
  \bibfield  {author} {\bibinfo {author} {\bibfnamefont {M.}~\bibnamefont
  {Aspelmeyer}}, \bibinfo {author} {\bibfnamefont {T.~J.}\ \bibnamefont
  {Kippenberg}}, \ and\ \bibinfo {author} {\bibfnamefont {F.}~\bibnamefont
  {Marquardt}},\ }Cavity optomechanics,\ \href {\doibase
  10.1103/RevModPhys.86.1391} {\bibfield  {journal} {\bibinfo  {journal} {Rev.
  Mod. Phys.}\ }\textbf {\bibinfo {volume} {86}},\ \bibinfo {pages} {1391}
  (\bibinfo {year} {2014}{\natexlab{b}})}\BibitemShut {NoStop}%
\bibitem [{\citenamefont {Bushev}\ \emph {et~al.}(2019)\citenamefont {Bushev},
  \citenamefont {Bourhill}, \citenamefont {Goryachev}, \citenamefont
  {Kukharchyk}, \citenamefont {Ivanov}, \citenamefont {Galliou}, \citenamefont
  {Tobar},\ and\ \citenamefont {Danilishin}}]{PhysRevD.100.066020}%
  \BibitemOpen
  \bibfield  {author} {\bibinfo {author} {\bibfnamefont {P.~A.}\ \bibnamefont
  {Bushev}}, \bibinfo {author} {\bibfnamefont {J.}~\bibnamefont {Bourhill}},
  \bibinfo {author} {\bibfnamefont {M.}~\bibnamefont {Goryachev}}, \bibinfo
  {author} {\bibfnamefont {N.}~\bibnamefont {Kukharchyk}}, \bibinfo {author}
  {\bibfnamefont {E.}~\bibnamefont {Ivanov}}, \bibinfo {author} {\bibfnamefont
  {S.}~\bibnamefont {Galliou}}, \bibinfo {author} {\bibfnamefont {M.~E.}\
  \bibnamefont {Tobar}}, \ and\ \bibinfo {author} {\bibfnamefont
  {S.}~\bibnamefont {Danilishin}},\ }Testing the generalized uncertainty
  principle with macroscopic mechanical oscillators and pendulums,\ \href
  {\doibase 10.1103/PhysRevD.100.066020} {\bibfield  {journal} {\bibinfo
  {journal} {Phys. Rev. D}\ }\textbf {\bibinfo {volume} {100}},\ \bibinfo
  {pages} {066020} (\bibinfo {year} {2019})}\BibitemShut {NoStop}%
\bibitem [{\citenamefont {Verhagen}\ \emph {et~al.}(2012)\citenamefont
  {Verhagen}, \citenamefont {Del{\'e}glise}, \citenamefont {Weis},
  \citenamefont {Schliesser},\ and\ \citenamefont
  {Kippenberg}}]{verhagen2012quantum}%
  \BibitemOpen
  \bibfield  {author} {\bibinfo {author} {\bibfnamefont {E.}~\bibnamefont
  {Verhagen}}, \bibinfo {author} {\bibfnamefont {S.}~\bibnamefont
  {Del{\'e}glise}}, \bibinfo {author} {\bibfnamefont {S.}~\bibnamefont {Weis}},
  \bibinfo {author} {\bibfnamefont {A.}~\bibnamefont {Schliesser}}, \ and\
  \bibinfo {author} {\bibfnamefont {T.~J.}\ \bibnamefont {Kippenberg}},\
  }Quantum-coherent coupling of a mechanical oscillator to an optical cavity
  mode,\ \href {\doibase 10.1038/nature10787} {\bibfield  {journal} {\bibinfo
  {journal} {Nature}\ }\textbf {\bibinfo {volume} {482}},\ \bibinfo {pages}
  {63} (\bibinfo {year} {2012})}\BibitemShut {NoStop}%
\bibitem [{\citenamefont {Song}\ \emph {et~al.}(2018)\citenamefont {Song},
  \citenamefont {Wang},\ and\ \citenamefont {Li}}]{SONG201839}%
  \BibitemOpen
  \bibfield  {author} {\bibinfo {author} {\bibfnamefont {L.}~\bibnamefont
  {Song}}, \bibinfo {author} {\bibfnamefont {Z.}~\bibnamefont {Wang}}, \ and\
  \bibinfo {author} {\bibfnamefont {Y.}~\bibnamefont {Li}},\ }Enhancing optical
  nonreciprocity by an atomic ensemble in two coupled cavities,\ \href
  {\doibase https://doi.org/10.1016/j.optcom.2018.01.009} {\bibfield  {journal}
  {\bibinfo  {journal} {Opt. Commun.}\ }\textbf {\bibinfo {volume} {415}},\
  \bibinfo {pages} {39} (\bibinfo {year} {2018})}\BibitemShut {NoStop}%
\bibitem [{\citenamefont {Vitali}\ and\ \citenamefont
  {Tombesi}(2007)}]{Vitali2007}%
  \BibitemOpen
  \bibfield  {author} {\bibinfo {author} {\bibfnamefont {D.}~\bibnamefont
  {Vitali}}\ and\ \bibinfo {author} {\bibfnamefont {P.}~\bibnamefont
  {Tombesi}},\ }Quantum entanglement between mechanical oscillators and the
  electromagnetic field in cavity optomechanics,\ \href {\doibase
  10.1103/PhysRevA.76.032105} {\bibfield  {journal} {\bibinfo  {journal} {Phys.
  Rev. A}\ }\textbf {\bibinfo {volume} {76}},\ \bibinfo {pages} {032105}
  (\bibinfo {year} {2007})}\BibitemShut {NoStop}%
\bibitem [{\citenamefont {Macovei}\ \emph {et~al.}(2010)\citenamefont
  {Macovei}, \citenamefont {Evers},\ and\ \citenamefont
  {Keitel}}]{Macovei2010QuantumEI}%
  \BibitemOpen
  \bibfield  {author} {\bibinfo {author} {\bibfnamefont {M.}~\bibnamefont
  {Macovei}}, \bibinfo {author} {\bibfnamefont {J.}~\bibnamefont {Evers}}, \
  and\ \bibinfo {author} {\bibfnamefont {C.~H.}\ \bibnamefont {Keitel}},\
  }Quantum entanglement in dense multiqubit systems,\ \href {\doibase
  10.1080/09500341003654443} {\bibfield  {journal} {\bibinfo  {journal} {J.
  Mod. Opt.}\ }\textbf {\bibinfo {volume} {57}},\ \bibinfo {pages} {1287}
  (\bibinfo {year} {2010})}\BibitemShut {NoStop}%
\bibitem [{\citenamefont {Brooks}\ \emph {et~al.}(2012)\citenamefont {Brooks},
  \citenamefont {Botter}, \citenamefont {Schreppler}, \citenamefont {Purdy},
  \citenamefont {Brahms},\ and\ \citenamefont {Stamper-Kurn}}]{1132}%
  \BibitemOpen
  \bibfield  {author} {\bibinfo {author} {\bibfnamefont {D.~W.~C.}\
  \bibnamefont {Brooks}}, \bibinfo {author} {\bibfnamefont {T.}~\bibnamefont
  {Botter}}, \bibinfo {author} {\bibfnamefont {S.}~\bibnamefont {Schreppler}},
  \bibinfo {author} {\bibfnamefont {T.~P.}\ \bibnamefont {Purdy}}, \bibinfo
  {author} {\bibfnamefont {N.}~\bibnamefont {Brahms}}, \ and\ \bibinfo {author}
  {\bibfnamefont {D.~M.}\ \bibnamefont {Stamper-Kurn}},\ }Non-classical light
  generated by quantum-noise-driven cavity optomechanics,\ \href {\doibase
  10.1038/nature11325} {\bibfield  {journal} {\bibinfo  {journal} {Nature}\
  }\textbf {\bibinfo {volume} {488}},\ \bibinfo {pages} {476} (\bibinfo {year}
  {2012})}\BibitemShut {NoStop}%
\bibitem [{\citenamefont {Kronwald}\ \emph {et~al.}(2014)\citenamefont
  {Kronwald}, \citenamefont {Marquardt},\ and\ \citenamefont
  {Clerk}}]{Kronwald2014}%
  \BibitemOpen
  \bibfield  {author} {\bibinfo {author} {\bibfnamefont {A.}~\bibnamefont
  {Kronwald}}, \bibinfo {author} {\bibfnamefont {F.}~\bibnamefont {Marquardt}},
  \ and\ \bibinfo {author} {\bibfnamefont {A.~A.}\ \bibnamefont {Clerk}},\
  }Dissipative optomechanical squeezing of light,\ \href {\doibase
  10.1088/1367-2630/16/6/063058} {\bibfield  {journal} {\bibinfo  {journal}
  {New J. Phys.}\ }\textbf {\bibinfo {volume} {16}},\ \bibinfo {pages} {063058}
  (\bibinfo {year} {2014})}\BibitemShut {NoStop}%
\bibitem [{\citenamefont {Pan}\ \emph {et~al.}(2020)\citenamefont {Pan},
  \citenamefont {Xiao},\ and\ \citenamefont {Gao}}]{Pan_2020}%
  \BibitemOpen
  \bibfield  {author} {\bibinfo {author} {\bibfnamefont {G.}~\bibnamefont
  {Pan}}, \bibinfo {author} {\bibfnamefont {R.}~\bibnamefont {Xiao}}, \ and\
  \bibinfo {author} {\bibfnamefont {J.}~\bibnamefont {Gao}},\ }Enhanced
  entanglement and output squeezing of optomechanical system via a single
  four-level atom,\ \href {\doibase 10.1088/1612-202X/ab9903} {\bibfield
  {journal} {\bibinfo  {journal} {Laser Phys. Lett.}\ }\textbf {\bibinfo
  {volume} {17}},\ \bibinfo {pages} {085204} (\bibinfo {year}
  {2020})}\BibitemShut {NoStop}%
\bibitem [{\citenamefont {Turek}\ \emph {et~al.}(2014)\citenamefont {Turek},
  \citenamefont {Yang}, \citenamefont {Maimaiti}, \citenamefont {Li},\ and\
  \citenamefont {Sun}}]{PhysRevA.90.013836}%
  \BibitemOpen
  \bibfield  {author} {\bibinfo {author} {\bibfnamefont {Y.}~\bibnamefont
  {Turek}}, \bibinfo {author} {\bibfnamefont {L.~P.}\ \bibnamefont {Yang}},
  \bibinfo {author} {\bibfnamefont {W.}~\bibnamefont {Maimaiti}}, \bibinfo
  {author} {\bibfnamefont {Y.}~\bibnamefont {Li}}, \ and\ \bibinfo {author}
  {\bibfnamefont {C.~P.}\ \bibnamefont {Sun}},\ }Indirect driving of a
  cavity-QED system and its induced nonlinearity,\ \href {\doibase
  10.1103/PhysRevA.90.013836} {\bibfield  {journal} {\bibinfo  {journal} {Phys.
  Rev. A}\ }\textbf {\bibinfo {volume} {90}},\ \bibinfo {pages} {013836}
  (\bibinfo {year} {2014})}\BibitemShut {NoStop}%
\bibitem [{\citenamefont {Sun}\ \emph {et~al.}(2003)\citenamefont {Sun},
  \citenamefont {Li},\ and\ \citenamefont {Liu}}]{PhysRevLett.91.147903}%
  \BibitemOpen
  \bibfield  {author} {\bibinfo {author} {\bibfnamefont {C.~P.}\ \bibnamefont
  {Sun}}, \bibinfo {author} {\bibfnamefont {Y.}~\bibnamefont {Li}}, \ and\
  \bibinfo {author} {\bibfnamefont {X.~F.}\ \bibnamefont {Liu}},\
  }Quasi-Spin-Wave Quantum Memories with a Dynamical Symmetry,\ \href {\doibase
  10.1103/PhysRevLett.91.147903} {\bibfield  {journal} {\bibinfo  {journal}
  {Phys. Rev. Lett.}\ }\textbf {\bibinfo {volume} {91}},\ \bibinfo {pages}
  {147903} (\bibinfo {year} {2003})}\BibitemShut {NoStop}%
\bibitem [{\citenamefont {Holstein}\ and\ \citenamefont
  {Primakoff}(1940)}]{PhysRev.58.1098}%
  \BibitemOpen
  \bibfield  {author} {\bibinfo {author} {\bibfnamefont {T.}~\bibnamefont
  {Holstein}}\ and\ \bibinfo {author} {\bibfnamefont {H.}~\bibnamefont
  {Primakoff}},\ }Field Dependence of the Intrinsic Domain Magnetization of a
  Ferromagnet,\ \href {\doibase 10.1103/PhysRev.58.1098} {\bibfield  {journal}
  {\bibinfo  {journal} {Phys. Rev.}\ }\textbf {\bibinfo {volume} {58}},\
  \bibinfo {pages} {1098} (\bibinfo {year} {1940})}\BibitemShut {NoStop}%
\bibitem [{\citenamefont {Jin}\ \emph {et~al.}(2003)\citenamefont {Jin},
  \citenamefont {Zhang}, \citenamefont {Liu},\ and\ \citenamefont
  {Sun}}]{PhysRevB.68.134301}%
  \BibitemOpen
  \bibfield  {author} {\bibinfo {author} {\bibfnamefont {G.~R.}\ \bibnamefont
  {Jin}}, \bibinfo {author} {\bibfnamefont {P.}~\bibnamefont {Zhang}}, \bibinfo
  {author} {\bibfnamefont {Y.-x.}\ \bibnamefont {Liu}}, \ and\ \bibinfo
  {author} {\bibfnamefont {C.~P.}\ \bibnamefont {Sun}},\ }Superradiance of
  low-density Frenkel excitons in a crystal slab of three-level atoms: The
  quantum interference effect,\ \href {\doibase 10.1103/PhysRevB.68.134301}
  {\bibfield  {journal} {\bibinfo  {journal} {Phys. Rev. B}\ }\textbf {\bibinfo
  {volume} {68}},\ \bibinfo {pages} {134301} (\bibinfo {year}
  {2003})}\BibitemShut {NoStop}%
\bibitem [{\citenamefont {Song}\ \emph {et~al.}(2005)\citenamefont {Song},
  \citenamefont {Zhang}, \citenamefont {Shi},\ and\ \citenamefont
  {Sun}}]{PhysRevB.71.205314}%
  \BibitemOpen
  \bibfield  {author} {\bibinfo {author} {\bibfnamefont {Z.}~\bibnamefont
  {Song}}, \bibinfo {author} {\bibfnamefont {P.}~\bibnamefont {Zhang}},
  \bibinfo {author} {\bibfnamefont {T.}~\bibnamefont {Shi}}, \ and\ \bibinfo
  {author} {\bibfnamefont {C.-P.}\ \bibnamefont {Sun}},\ }Effective boson-spin
  model for nuclei-ensemble-based universal quantum memory,\ \href {\doibase
  10.1103/PhysRevB.71.205314} {\bibfield  {journal} {\bibinfo  {journal} {Phys.
  Rev. B}\ }\textbf {\bibinfo {volume} {71}},\ \bibinfo {pages} {205314}
  (\bibinfo {year} {2005})}\BibitemShut {NoStop}%
\bibitem [{\citenamefont {Giovannetti}\ and\ \citenamefont
  {Vitali}(2001)}]{PhysRevA.63.023812}%
  \BibitemOpen
  \bibfield  {author} {\bibinfo {author} {\bibfnamefont {V.}~\bibnamefont
  {Giovannetti}}\ and\ \bibinfo {author} {\bibfnamefont {D.}~\bibnamefont
  {Vitali}},\ }Phase-noise measurement in a cavity with a movable mirror
  undergoing quantum Brownian motion,\ \href {\doibase
  10.1103/PhysRevA.63.023812} {\bibfield  {journal} {\bibinfo  {journal} {Phys.
  Rev. A}\ }\textbf {\bibinfo {volume} {63}},\ \bibinfo {pages} {023812}
  (\bibinfo {year} {2001})}\BibitemShut {NoStop}%
\bibitem [{\citenamefont {Nie}\ \emph {et~al.}(2016)\citenamefont {Nie},
  \citenamefont {Chen},\ and\ \citenamefont {Lan}}]{PhysRevA.93.023841}%
  \BibitemOpen
  \bibfield  {author} {\bibinfo {author} {\bibfnamefont {W.}~\bibnamefont
  {Nie}}, \bibinfo {author} {\bibfnamefont {A.}~\bibnamefont {Chen}}, \ and\
  \bibinfo {author} {\bibfnamefont {Y.}~\bibnamefont {Lan}},\ }Optical-response
  properties in levitated optomechanical systems beyond the low-excitation
  limit,\ \href {\doibase 10.1103/PhysRevA.93.023841} {\bibfield  {journal}
  {\bibinfo  {journal} {Phys. Rev. A}\ }\textbf {\bibinfo {volume} {93}},\
  \bibinfo {pages} {023841} (\bibinfo {year} {2016})}\BibitemShut {NoStop}%
\bibitem [{\citenamefont {Hilico}\ \emph {et~al.}(1992)\citenamefont {Hilico},
  \citenamefont {Fabre}, \citenamefont {Reynaud},\ and\ \citenamefont
  {Giacobino}}]{PhysRevA.46.4397}%
  \BibitemOpen
  \bibfield  {author} {\bibinfo {author} {\bibfnamefont {L.}~\bibnamefont
  {Hilico}}, \bibinfo {author} {\bibfnamefont {C.}~\bibnamefont {Fabre}},
  \bibinfo {author} {\bibfnamefont {S.}~\bibnamefont {Reynaud}}, \ and\
  \bibinfo {author} {\bibfnamefont {E.}~\bibnamefont {Giacobino}},\ }Linear
  input-output method for quantum fluctuations in optical bistability with
  two-level atoms,\ \href {\doibase 10.1103/PhysRevA.46.4397} {\bibfield
  {journal} {\bibinfo  {journal} {Phys. Rev. A}\ }\textbf {\bibinfo {volume}
  {46}},\ \bibinfo {pages} {4397} (\bibinfo {year} {1992})}\BibitemShut
  {NoStop}%
\bibitem [{\citenamefont {Nie}\ \emph {et~al.}(2012)\citenamefont {Nie},
  \citenamefont {Lan}, \citenamefont {Li},\ and\ \citenamefont
  {Zhu}}]{PhysRevA.86.063809}%
  \BibitemOpen
  \bibfield  {author} {\bibinfo {author} {\bibfnamefont {W.}~\bibnamefont
  {Nie}}, \bibinfo {author} {\bibfnamefont {Y.}~\bibnamefont {Lan}}, \bibinfo
  {author} {\bibfnamefont {Y.}~\bibnamefont {Li}}, \ and\ \bibinfo {author}
  {\bibfnamefont {S.}~\bibnamefont {Zhu}},\ }Effect of the Casimir force on the
  entanglement between a levitated nanosphere and cavity modes,\ \href
  {\doibase 10.1103/PhysRevA.86.063809} {\bibfield  {journal} {\bibinfo
  {journal} {Phys. Rev. A}\ }\textbf {\bibinfo {volume} {86}},\ \bibinfo
  {pages} {063809} (\bibinfo {year} {2012})}\BibitemShut {NoStop}%
\bibitem [{\citenamefont {DeJesus}\ and\ \citenamefont
  {Kaufman}(1987)}]{PhysRevA.35.5288}%
  \BibitemOpen
  \bibfield  {author} {\bibinfo {author} {\bibfnamefont {E.~X.}\ \bibnamefont
  {DeJesus}}\ and\ \bibinfo {author} {\bibfnamefont {C.}~\bibnamefont
  {Kaufman}},\ }Routh-Hurwitz criterion in the examination of eigenvalues of a
  system of nonlinear ordinary differential equations,\ \href {\doibase
  10.1103/PhysRevA.35.5288} {\bibfield  {journal} {\bibinfo  {journal} {Phys.
  Rev. A}\ }\textbf {\bibinfo {volume} {35}},\ \bibinfo {pages} {5288}
  (\bibinfo {year} {1987})}\BibitemShut {NoStop}%
\bibitem [{\citenamefont {Chen}\ \emph
  {et~al.}(2017{\natexlab{b}})\citenamefont {Chen}, \citenamefont {Lin},
  \citenamefont {He},\ and\ \citenamefont {Lin}}]{Chen:17}%
  \BibitemOpen
  \bibfield  {author} {\bibinfo {author} {\bibfnamefont {Z.~X.}\ \bibnamefont
  {Chen}}, \bibinfo {author} {\bibfnamefont {Q.}~\bibnamefont {Lin}}, \bibinfo
  {author} {\bibfnamefont {B.}~\bibnamefont {He}}, \ and\ \bibinfo {author}
  {\bibfnamefont {Z.~Y.}\ \bibnamefont {Lin}},\ }Entanglement dynamics in
  double-cavity optomechanical systems,\ \href {\doibase 10.1364/OE.25.017237}
  {\bibfield  {journal} {\bibinfo  {journal} {Opt. Express}\ }\textbf {\bibinfo
  {volume} {25}},\ \bibinfo {pages} {17237} (\bibinfo {year}
  {2017}{\natexlab{b}})}\BibitemShut {NoStop}%
\bibitem [{\citenamefont {Zhang}\ \emph {et~al.}(2013)\citenamefont {Zhang},
  \citenamefont {Zhou},\ and\ \citenamefont {Zhou}}]{Zhang_2013}%
  \BibitemOpen
  \bibfield  {author} {\bibinfo {author} {\bibfnamefont {X.-F.}\ \bibnamefont
  {Zhang}}, \bibinfo {author} {\bibfnamefont {Y.}~\bibnamefont {Zhou}}, \ and\
  \bibinfo {author} {\bibfnamefont {L.}~\bibnamefont {Zhou}},\ }Enhancing
  stationary optomechanical entanglement with the Kerr medium,\ \href {\doibase
  10.1088/1674-1056/22/6/064206} {\bibfield  {journal} {\bibinfo  {journal}
  {Chin. Phys. B}\ }\textbf {\bibinfo {volume} {22}},\ \bibinfo {pages}
  {064206} (\bibinfo {year} {2013})}\BibitemShut {NoStop}%
\end{thebibliography}%

\end{document}